\documentclass[10pt,aps,amsfonts,pra,twocolumn,showpacs]{revtex4-1}
\usepackage{verbatim,epsfig,amsmath,amssymb,bm,epsf,graphicx,psfrag,bbold,amsthm,amsfonts}
\usepackage{hyperref}
\usepackage[all]{xy}
\usepackage{color}

\newcommand{\ket}[1]{|#1\rangle}
\newcommand{\bra}[1]{\langle #1|}
\newcommand{\innerproduct}[2]{\langle #1| #2\rangle}

\begin{document}

\title{Global and short-range entanglement properties in excited, many-body localized spin chains}

\author{Colin G. West}
\affiliation{Department of Physics, University of Colorado at Boulder, Department of Physics, University of California at Santa Cruz, Santa Cruz, CA 95064, USA}
\affiliation{C. N. Yang Institute for Theoretical Physics, Stony Brook, NY 11794-3840, USA}

\author{Tzu-Chieh Wei}
\affiliation{C. N. Yang Institute for Theoretical Physics and
	Department of Physics and Astronomy, State University of New York at
	Stony Brook, Stony Brook, NY 11794-3840, USA}

\date{\today}

\begin{abstract}
We explore the use of short-range entanglement measures, such 
as concurrence and negativity, and global entanglement measures such as 
geometric entanglement, as indicators of many-body localization (MBL) in 
the spectra of disordered spin systems. From the perspective of 
entanglement monogamy, the two types of entanglement behave oppositely 
in the thermalized and MBL phases. In a recent work, the concurrence of 
subsystems, a measure of local entanglement, was used in a study of many-body localization in a 
one-dimensional spin-$1/2$ system~\cite{Bera}. We show numerically that the 
negativity displays notably similar behavior for this system, with the 
advantage that it can also be extended to systems of higher local 
dimension. We then demonstrate this extension in practice by using it to 
predict the existence of an MBL phase in a disordered a spin-1 system. 
In terms of global entanglement, the geometric entanglement of both 
spin-$1/2$ and spin-1 systems is also shown to behave as a complementary 
indicator of the MBL phenomenon.

\end{abstract}

 \maketitle
 
 \section{Introduction}
 
Because ergodicity is a central assumption of traditional statistical and thermal physics, exceptions to ergodic behavior in a thermodynamic system have long been of particular interest. Typically, the notion of ergodicity is formalized through the so-called ``eigenstate thermalization hypothesis" (ETH)\cite{JoshThermalization, SrednickiThermalization, SrednickiThermalization2, ETH3, ETH4}. The ETH (reviewed below), predicts thermal behavior in each individual eigenstate if a system in general is to be described as ``thermal." It is clear that this condition can be violated in integrable systems, but it is beginning to be shown that such violations can also be found under conditions which appear to be far more generic.

 Anderson's groundbreaking work~\cite{AndersonOriginal} explained clearly how a single particle in a sufficiently disordered medium can see its wavefunction become spatially localized, and also speculated that this kind of localization could persist in many-particle regimes, if the strength of the interactions is sufficiently weak compared to the disorder.  In the past decade, theoretical and numerical evidence~\cite{Prosen, PalOriginal, Huse, HuseMBL, MBLEntGrowth, LBits1,BAA, HuseFermions} has begun to strongly support this conjecture, demonstrating the phenomenon which has come to be known as ``many body localization" (MBL). More recently, direct experimental observation of this ergodicity-breaking has emerged as well~\cite{MBLExperiment}.

In the current study of the phenomenon, one important goal is to be able to predict the points at which a system will transition from ergodic to localized behavior, as the strength of the disorder is increased. This transition point is the so-called ``critical value" of the disorder, as it can be thought of as marking the boundary between a thermal, delocalized phase, and a more exotic localized phase.  A wide variety of quantities have been studied as indicators to detect MBL behavior in a system, and identify the critical value of the disorder. Such quantities include the behavior of local observables, the level statistics, and the participation ratio\cite{Huse, Alet} of particular states. Another class of MBL indicators are quantities which measure, in some sense, the distribution of entanglement within a state. In a localized system, one expects various localized ``pockets" of the system to be strongly entangled within themselves, and from the monogamy of entanglement we thus also expect them to be only weakly entangled with the rest of the system. By contrast, in a thermalized state, every choice of subsystem sees the remainder of the state as a thermal reservoir; one consequently expects considerable entanglement across all subsystems in the thermal case~\cite{PalOriginal, HuseMBL}. Thus, from the monogamy of entanglement, the entanglement between two small subsystems is also expected to be correspondingly small.

In this paper we consider three different measures of entanglement as indicators of MBL behavior. The first, ``concurrence" \cite{WoottersCCROriginal1, WoottersCCROriginal2}, is an entanglement monotone which is related to entanglement of formation, but which can be more readily computed for certain cases of mixed states, notably for including two-qubit systems. As such, it was recently studied in the context of MBL phenomena\cite{Bera} as a way of quantifying entanglement between nearest-neighbor particles in a spin-1/2 system. In that work, it is shown that the apparent transition point between ergodic and localized phases coincides with a rapid increase in entanglement between nearest neighbors, as measured by concurrence, signaling the fact that various subsystems have begun to emerge which are relatively isolated from their surroundings. 

However, as originally proposed, a closed-form formula for concurrence exists only for the case of quibit-qubit systems\cite{WoottersCCRFormula}. Although various generalizations and extensions have been proposed \cite{CCRGeneralization, CCRGeneralization2}, for generic mixed states of higher spin it remains difficult to calculate. For this reason, we consider also the so-called ``negativity" of nearest-neighbor subsystems. The negativity (explained in detail below) is a similar measure of entanglement whose relationship to concurrence is precisely known in the case of quibit-quibit systems\cite{NegVsCCR}, but which can also be computed for systems of arbitrary dimension, making it a strong candidate to be an MBL indicator in systems with higher local dimension.

In addition to these local measures of entanglement, which we use to consider the entanglement of nearest-neighbors, we consider the complemantary concept of geometric entanglement, which measures the global entanglement of the entire state\cite{WeiGeometricOriginal, WeiGeometric2}. Although we expect its behavior to be opposite that of the other entanglement measures (decreasing suddenly as one moves into the localized regime), it is also useful as an identifier of the MBL transition, as we shall demonstrate.
  
 This paper is organized as follows: in sections \ref{Background} and \ref{Indicators}, we review the nature of the MBL phenomenon, and the three measures of entanglement considered here (concurrence, negativity, and geometric entanglement). In section \ref{Algorithms}, we briefly describe the nature of the numerical methods used in generating our results. Section \ref{SpinHalf} demonstrates the use of these entanglement measures as applied to a disordered spin 1/2 system, and section \ref{Spin1} shows the extension of these measures to a system of higher spin (spin 1). A summary of this work is presented in section \ref{Summary}.

 \section{Entanglement and MBL}\label{Background}
 
In this section, we review as background the ETH and its role in understanding the MBL phenomenon. We then review the entanglement measures which will be used in this paper, and their connections to the known properties of MBL systems.
 
 \subsection{MBL in quantum spin systems}
 
Generally speaking, the expected behavior of a closed, interacting quantum system $S$ is that, whatever its initial configuration, with time the system will ``thermalize" to an equilibrium state. Although the information about the system's initial condition is not ``lost" (it cannot be, since the time evolution of a quantum system is a unitary process), with time the information becomes locally unobservable, having decohered across the entire state.  For long timescales, the system is expected to reach a ``thermalized" state, in which it resembles the state that the system would be in if it were in contact with a thermal reservoir. 

This word ``resembles" can be made more quantitative as follows. If it \emph{were} in contact with a thermal reservoir at a temperature $T$, the state of the system $S$ would be described by a thermal density matrix

\begin{equation}\label{eq:ThermalDensityMatrix}
\rho_{th}(T) = \frac{1}{Z}e^{-H/k_bT}
\end{equation}
where $Z$ is the standard partition function. It follows that in such a case, we we could also consider the state of an arbitrary subsystem $A \subset S$ by tracing over the environment $E = S \setminus A$, i.e.

\begin{equation}\label{eq:ThermalDensityMatrixTraced}
\rho_{th}^{(A)}(T) = tr_E \left( \frac{1}{Z}e^{-H/k_bT} \right).
\end{equation}
 
We emphasize again that Eq. \ref{eq:ThermalDensityMatrix} is \emph{not} the actual state of the system when it has thremalized. It cannot have wound up in a general Boltzmann distribution, which contains no information about its initial state, because it evolves only through unitary (information-preserving) evolution. By contrast, Eq. \ref{eq:ThermalDensityMatrixTraced} \emph{can} be the correct description of some subsystem $A$, because the remainder of the system, $E$, appears as a reservoir from the perspective of $A$ in the thermodynamic limit. When we say a state has thermalized if it ``resembles" a state in contact with a thermal reservoir (Eq. \ref{eq:ThermalDensityMatrix}), we mean more precisely that for any choice of subsystem $A$, the state of that subsystem is described by Eq. \ref{eq:ThermalDensityMatrixTraced} \cite{JoshThermalization, SrednickiThermalization}.
 
In a thermalizing system, we consequently expect a type of behavior which has come to be known as the ``eigenstate thermalization hypothesis" (ETH)\cite{JoshThermalization, SrednickiThermalization, SrednickiThermalization2, ETH3, ETH4}. Suppose the system had been initialized in any exact eigenstate of the governing Hamiltonian: in such case, naturally the state of the system will not change during its time evolution. But we know that, for a thermalizing system, after long time periods its subsystems can all be described by Eq. \ref{eq:ThermalDensityMatrixTraced} for long timescales). Consequently, the ETH states that in a thermal system, all eigenstates must by themselves \emph{already} be thermalized states.

Many-body localization (MBL) can be thought of as a phenomenon in disordered quantum systems in which the ETH is explicitly false: states exist (both eigenstates and general linear combinations thereof) which are not thermal in the formal sense defined above. The behavior of such states defy our classical intuition: in a system with interactions, which one might reasonably presume would allow every subsystem to ``talk" to every other, one might imagine that all initial information would becomes distributed across the state. But instead, such systems display a kind of ``memory" of their initial conditions. Even more strikingly, in an MBL system, this behavior is stable against perturbations in the disorder, unlike the comparable behavior of a finely-tuned integrable system, which is not.

These properties can be viewed more formally through the notion of an ``l-bit" (for ``localized bit")~\cite{LBits1, LBits2}. Consider a system of noninteracting particles; such a system would naturally be described by a set of conserved charges in some basis. For example, in a trivial, non-interacting spin chain $H = \sum_i \sigma^z_i$, eigenstates can all be described by the set of local spins $\{ \sigma^z_i \}$, which all commute with the Hamiltonian and hence provide good quantum numbers for a basis of simultaneous eigenstates. But if this behavior exists for isolated spins (``bits") which are \emph{explicitly} not interacting, then the same behavior should hold for localized systems as well, so that a new family of conserved charges (the  l-bits) emerges. In particular, it has been shown\cite{LBits1} that these l-bits are simply ``dressed" versions of the pre-existing bits: they are given by products (and sums of products) of the original bits at nearby sites. Contributions from sites which are far away are exponentially suppressed, giving rise to a new set of quantum numbers for the system. This behavior is the reason that measures of global and local entanglement are well-suited to detecting MBL behavior, as discussed in the next section, and will also form the basis for the MPS algorithms used for computing MBL states (discussed in section~\ref{Algorithms}). 

\section{Entanglement Indicators of Localization}\label{Indicators}

Qualitatively speaking, it is perhaps intuitive that as a system displays localized behavior, we find certain subsets which are internally highly entangled, while being only weakly entangled with the rest of the system. To this end, Ref.~\cite{Bera} proposes the use of a local measure of entanglement as an indicator of MBL, in the following way. In a localized state, certain, though not all, local  subsystems are expected to be highly entangled. Hence, for example, in a one-dimensional system, if one computes the entanglement between pairs of nearest-neighbors (tracing out the remainder of the system) and takes the sum or average over all such pairs, the resulting quantity is expected to be large for a localized system and small for a thermalized one. Note that the entanglement measure used must be suitable for use in a mixed state, since we will be dealing always with subsystems after tracing out the rest of the state. A mixed state is unentangled if it can be expressed as a mixture of unentangled pure states; otherwise it is an entangled state~\cite{EntanglementReview}. We now discuss two possible measures of entanglement appropriate for quantifying entanglement in such a setting.
 \subsection{Concurrence}
Concurrence first arose as part of an important formula for the entanglement of formation~\cite{WoottersCCRFormula}, but has emerged as an important entanglement measure in its own right. Originally, concurrence was defined for two-qubit systems (e.g. two spin-1/2 systems or their equivalent), and although generalizations to higher dimensional systems have been proposed~\cite{WoottersCCROriginal2, CCRGeneralization, CCRGeneralization2}, this remains the only setting in which a convenient and closed-form expression has been obtained which would allow for direct calculation in mixed states. In this context, the concurrence is given by comparing the two-qubit system with a spin-flipped version of itself. If the system were a pure state in the computational ($\sigma^z$) basis, we could effectuate the spin-flip with a time-reversal operation, i.e. with complex conjugation and the Pauli operator $\sigma^y$ on each spin. Hence if we have instead a mixed state described by a density matrix $\rho$, its spin-flipped counterpart is given by

\begin{equation}
\tilde{\rho} = (\sigma^y \otimes \sigma^y)\rho^*(\sigma^y \otimes \sigma^y)
\end{equation}

As shown in~\cite{WoottersCCROriginal2}, we can now construct a sensible measure of entanglement by taking the matrix $\rho \tilde{\rho}$ and finding its eigenvalues $\lambda_i$. Let these eigenvalues be in descending order so that $\lambda_1 \geq \lambda2 \geq \dots$. In this case, the concurrence is given by

\begin{equation}\label{eq:Concurrence}
\mathcal{C} = \max( 0, \sqrt{\lambda_1} - \sqrt{\lambda_2} - \sqrt{\lambda_3} - \sqrt{\lambda_4} ). 
\end{equation}
Note that the eigenvalues of $\rho \tilde{\rho}$ are guaranteed to be nonnegative so that Eq.~\ref{eq:Concurrence} is always well-defined.

As a measure of entanglement in mixed states, concurrence is potentially well-suited for use in the context of MBL systems, which are defined in terms of the properties of their (potentially mixed) subsystems. In particular, concurrence obeys monogamy of entanglement~\cite{MonogCCR}, so that when two spins are highly entangled with one another, they are less able to be entangled with the remainder of the system. Intuitively, this suggests that the concurrence between locally neighboring spins should be closely related to localization within the system; if subsystems are to be mutually connected in a manner that allows their environment to act as a reservoir, they cannot be too strongly entangled as pairs of particles. 

This observation motivated the authors of Ref.~\cite{Bera} to consider ``total nearest neighbor concurrence" $\mathcal{C}^{tot}_{NN}$ of a spin system as an indicator of localization, calculated as a sum over the concurrence of all nearest-neighbor pairs in the system

\begin{equation}\label{eq:CTot}
\mathcal{C}^{tot}_{NN} = \sum_{<ij>} \mathcal{C}^{ij}) 
\end{equation}

The resulting quantity should be small in a delocalized system, since few if any pairs of spins have large amounts of entanglement with each other, preferring instead to entangle with the broader environment. Conversely, the presence of even a few highly entangled pairs of neighboring spins, as in a localized state, can make the quantity grow rapidly. This behavior was demonstrated to coincide with other known indicators of localization in the context of a random, spin-1/2 Heisenberg model \cite{Bera}, discussed in greater detail below in section \ref{SpinHalf}.

 \subsection{Negativity}
 
 Concurrence is a powerful measure of entanglement for two-qubit systems. But as previously remarked, it also possesses substantial limitations which prevent it from addressing the full range of questions which arise in MBL systems. But an alternate measure of entanglement in mixed systems, the \emph{negativity} can be computed directly for subsystems of arbitrary dimension (not simply qubits) and arbitrary system sizes sizes (not simply two spins). It has also recently been shown that tensor network techniques can be used to extract negativity even for large and difficult systems~\cite{CalabreseNeg}, although within the scope of this work it will suffice to perform the straightforward calculation. 
 
 Negativity~\cite{NegativityOriginal} arose in the wake of the Peres-Horodecki~\cite{Peres, Horodecki} criterion, a necessary condition for a density matrix to represent a separable (unentangled) quantum state. This criterion states that, if $\rho$ is a density matrix on a composite space $\mathcal{H} = \mathcal{H}_A \otimes \mathcal{H}_B$, then $\rho$ is separable only if the partial transpose $\rho^{T_A}$ has non-negative eigenvalues. In our case, where we are particularly interested in measuring entanglement, it is the contrapositive of this statement which holds greater significance: if $\rho^{T_A}$ \emph{does} have negative eigenvalues, $\rho$ must have represented an entangled state (with respect to the systems $\mathcal{H}_A$ and $\mathcal{H}_B$.
 
 Thus, we define the negativity as a measure of entanglement to represent the extent to which this criterion is violated: it is the sum of the magnitudes of all negative eigenvalues which appear after partial transposition. Because a proper density matrix has unit trace, the size of this sum is inherently normalized and can be compared between systems. In practice, this recipe for the negativity can be compactly written as

\begin{equation}
\mathcal{N}(\rho, A) = \sum_i \frac{|\lambda_i| - \lambda_i}{2}
\end{equation}
where the elements $\{ \lambda_i \}$ represent the eigenvalues of $\rho^{T_A}$.

Since the negativity shares many properties with the concurrence, including notably the monogamy of entanglement~\cite{MonogGeneral}, it is natural to ask whether it also serves as an indicator of MBL, and if so, whether its use for this purpose can be extended to systems in which calculation of concurrence would be intractable, such as for the case of a spin-1 system. By analogue to how we approached the concurrence, we will consider the total nearest-neighbor negativity of a state. For each pair of nearest neighbors, we construct the (reduced) density matrix for these two spins, and then partially transpose with respect to one of the spins to compute the negativity. The sum of these negativities for all neighboring pairs give us our desired quantity

\begin{equation}\label{eq:NegNN}
 \mathcal{N}^{tot}_{NN} = \sum_{<ij>} \mathcal{N}^{ij}. 
\end{equation}

 \subsection{Geometric Entanglement}
 Finally, in contrast to the approach discussed above, using either the total concurrence or total negativity of nearest-neighbor pairs, one might also seek to detect localization with a measure of the \emph{global} entanglement of the state. A global entanglement measure would display the opposite behavior: it will be large for a thermalized state, with correlations distributed across the entire system, and smaller in a localized state where isolated pockets are more locally entangled. One of the first proposed global measures is the so-called ``relative entropy of entanglement"~\cite{WeiGeometric2}. This, however, is in general difficult to calculate, even for pure states. We turn instead to the concept of geometric entanglement, which can serve as a bound for the relative entropy of entanglemtns~\cite{WeiEricsson2004, hayashi2006}. For our application to MBL, we will compute the geometric entanglement for pure states (the individual many-body eigenstates), but we note that the concept has been extended to multipartite mixed states as well\cite{Geometric1, Geometric2, WeiGeometricOriginal}. 
 
Conceptually, the geometric entanglement of some $N$-body pure state $| \Psi \rangle$ is the distance in Hilbert space between $| \Psi \rangle$ and the nearest unentangled (product) state $|\Phi \rangle = \prod_{i = 1}^N | \phi_i \rangle$, which one might picture either as a distance or equivalently as an angle between the state vectors. Intuitively, when a state is only weakly entangled, and can be closely approximated by some product state, this distance is small. Of course, for a highly entangled state, the opposite is true. 

To capture this concept quantitatively, it is convenient to look at the square of the overlap

\begin{equation}
\Lambda(\Psi) =  \max _{| \Phi \rangle} |\langle \Psi| \Phi \rangle|^2 
\end{equation}
which naturally is directly related to measures of distance and can be more straightforward to calculate. Taking the logarithm produces a quantity which vanishes for unentangled states; further introducing a minus sign ensures that the quantity will be large when the entanglement is large. Hence the geometric entanglement of a state can be quantified as

\begin{equation}\label{eq:GeomEnt}
S_G = - \log \Lambda
\end{equation}

Particular details on computing this quantity numerically will be given in Sec.~\ref{sec:GeomEntMPS}. For the case of a mixed state $\rho$, this concept can be extended by taking the convex hull, minimizing over pure-state decompositions of $\rho$ \cite{WeiGeometricOriginal}, or alternatively as generalized in Ref.~\cite{hayashi2006}.
 
 \section{MBL States with Matrix Product Algorithms}\label{Algorithms}
 
Having identified three measures of entanglement which we wish to explore in the context of MBL phenomena (one of them, concurrence, having already been studied for this purpose in \cite{Bera}), we will now briefly review procedures for generating numerical representations of states, to which these measures can be applied. Because we wish to look for localization behavior in the context of disordered Hamiltonians, we will need to be able to generate a large number of such states, across a large number of disordered instances, so that we can average over disorder. For short chains, exact diagonalization using Lancsoz methods \cite{Lanczos1950, ImplicitLanczos} is by far the most effective means to generate these states, as the entire spectrum of a particular instance of a disordered Hamiltonian can be computed. With reasonably high-end computing resources, in a spin-1/2 system this is generally possible for chains up to a length of 16, or with considerably more run time per disordered instance, length 18 \cite{Bera}. Exact diagonalization techniques targeting a particular range of energies (but not the entire spectrum) can push this limit slightly higher \cite{Alet}. But the situation is much worse for the case of a spin-1 system, where the increase in the local dimension has an exponential impact on complexity of the problem. The same resources which make the spectrum of a 18-site Hamiltonian computable in a spin-1/2 case will be limited to systems of only eight or ten sites in the spin-1 case. 

Because many-body localization is fundamentally a phenomenon of the thermodynamic limit, it is important to push the frontier of system size when identifying and applying indicators of localization. For example, a common goal is to use these indicators to determine the critical value of disorder strength required to produce localization in a system. But finite-size effects tend to strongly increase the appearance of localization, meaning that when using only very short chains, one will tend to calculate a critical disorder which is smaller than it may be in the thermodynamic limit \cite{MBLFiniteEffects}.

To that end, we have employed a combination of exact diagonalization methods and algorithms based on ``Matrix Product State" (MPS) \cite{fannes1992, OriginalMPS3, MPS_Perez-Garcia2007} representations of the eigenstates. We now give a brief overview of the MPS formalism, particularly as it applies to the algorithms used in this work. For a more comprehensive review, we refer the reader to Refs.~\cite{TN1, MPSReview, MPS_Perez-Garcia2007}, which contain substantial technical details, and to Refs.~\cite{OrusReview, SydneyReview}, which are more pedagogically designed for introductory reading. 

 \subsection{Matrix Product States}
 
  Consider a one-dimensional spin chain of length $L$ and local spin $s$. The relevant Hilbert space for the entire system is a product of $L$ $d$-dimensional spaces, where $d = 2s-1$ is the dimension of the local Hilbert space associated with each individual spin. Naturally, the total Hilbert space is $d^L$ dimensional, and a general state vector for the spin chain is of the form
 
 \begin{equation} \label{eq:vector}
 \ket{\psi} = \sum_{i_1=1}^d \ldots \sum_{i_L=1}^d c_{i_1 \ldots i_L} \ket{i_1 \ldots i_L}
 \end{equation}
 for some tensor of coefficients $c_{i_1 \ldots i_L}$. The indices $i_1$, $i_2$, $i_L$ label the state of spins at each site. 
 
 As either $L$ or $d$ increase, the coefficient tensor quickly becomes unmanageably large, and cannot be represented numerically in its entirety. To avoid this ``catastrophe of dimensionality," The MPS approach seeks instead to represent the states of interest in the following form, with the massive single tensor $c_{i_1 \ldots i_L}$ replaced by a product of matrices $A^i$, with one set of matrices associated with each site. Hence at site $n$, we have a set of matrices $A^{i_n}$, where the index $i_n$ is the same as the spin-index of the site. For example, a state with generic open boundary conditions could be represented in the form

 \begin{equation} \label{eq:mps_open}
 \ket{\psi} = \sum_{i_1 \dots i_L}  \bra{v_{\text{left}}}  A^{i_1}_1 A^{i_2}_2 \dots  A^{i_L}_L \ket{v_{\text{right}}} \ket{i_1 \ldots i_L} 
 \end{equation}
 where $v_{\text{left}}$ and $v_{\text{right}}$ are vectors specifying boundary conditions, and where the subscripts ``$1$, $2$," etc., serve as a reminder that the set of matrices for site $1$ may be different from the set of matrices for site $2$, and so on.
 
 In general, a state in the form of Eq.~\ref{eq:vector} can also be represented in the form of Eq.~\ref{eq:mps_open} to arbitrary accuracy, provided that the dimension of each $D \times D$ matrix $A^i_m$ is allowed to become arbitrarily large, because the sets of MPS matrices $\left\{ A^{i_1} , A^{i_2}  \dots A^{i_L} \right\}$ can be constructed by successive singular value decompositions of the coefficient tensor $c_{i_1 \ldots i_N}$~\cite{vidal03}. But if the maximum dimension $D$ of these matrices (often called the ``bond dimension") can be arbitrarily large, it defeats the purpose. From a numerical perspective, the useful class of matrix product states are those where the maximum bond dimension can be kept to an relatively small finite limit.

 It has been known for some time that ground states of gapped, local Hamiltonians fall into this class of states, because the necessary bond dimension for such states approaches a constant independent of the length of the system~\cite{vidal03,hastings2007area}. This beneficial property occurs because the entanglement in such states obeys an ``area law," in the sense that the entanglement of any subsystem with its environment scales not with the volume of the subsystem but rather with the area of the boundary~\cite{hastings2007area, arad_itai}. This behavior is in marked contrast to the entanglement in a general eigenstate of an arbitrary Hamiltonian, but quite interestingly, it has recently also been realized that when a system displays many-body localization, \emph{almost all eigenstates obey an area law} with respect to the boundary of the subsystem~\cite{MBLAreaLaw1, MBLAreaLaw2, MBLisMPS}. Consequently, it is possible to efficiently represent as an MPS eigenstates of a localized system which would otherwise be much too large to consider computationally~\cite{MBLisMPS}. 
 
By analogy to Eq.~\ref{eq:mps_open}, one can also specify an \emph{operator} with coefficients given by a product of matrices: a so-called ``matrix product operator," or MPO~\cite{DMRGVerstraete2004, IanMPO2, IanMPO, DMRGMPSReview}, which has the form:
 
 \begin{equation}\label{eq:OBCMPO}
O = \sum_{s, s'} O^{s_1, s_2 \dots s_L}_{s'_1, s'_2, \dots s'_L}  \ket{s_1 s_2 \dots s_L}\bra{s'_1 s'_2 \dots s'_L}.
 \end{equation}
 with the coefficient tensor 
 \begin{equation}\label{eq:OBCMPO2}
O^{s_1, s_2 \dots s_L}_{s'_1, s'_2, \dots s'_L}  = \bra{v_{\text{left}}} M_{[1]}^{s_1, s'_1}M_{[2]}^{s_2 s'_2} \dots M_{[L]}^{s_L s'_L}\ket{v_{\text{right}}} 
\end{equation}
in which the coefficients of the operator are also represented through a product of matrices.

Like an MPS, an MPO description is particularly compact and efficient for certain classes of operators; particularly those which can be expressed as a sum of local operators~\cite{IanMPO, IanInfiniteBinder1}. This property creates an additional important restriction on our ability to study MBL states through matrix product algorithms: MBL states can be efficiently represented as an MPS, but in order to efficiently compute these states in the first place we will desire an algorithm that also does not require any inefficient operator representations. The SIMPS and DMRG-X algorithms of Ref.~\cite{SIMPS} cleverly satisfy this requirement; in this work we will use the ``SIMPS" algorithm, which is outlined below.
  
 \subsection{Generating MBL states with SIMPS}
 
 A number of different algorithms for constructing MPS representations of MBL states have been proposed~\cite{SIMPS, EnDMRG, SQUIMPS}; we briefly review here the ``Shift and Invert MPS" algorithm (``SIMPS")~\cite{SIMPS}, which we have used for this purpose in this work. 
 
 Broadly, the goal of the SIMPS algorithm (discussed here only for open-boundary systems) is as follows: assume we have a Hamiltonian and we wish to search for an MBL state with energy $\tilde{\lambda}$ close to a target value $\lambda$. Construct a shifted operator $O = H - \lambda$, and observe that the state $\ket{\psi_{\tilde{\lambda}}}$ with energy closest to our target can be found as the dominant eigenvector of the inverted Hamiltonian

\begin{equation}\label{eq:SIMPSeq}
\tilde{H} = (H-\lambda)^{-1} = O^{-1}.
\end{equation} 
 
To find this dominant eigenvector, SIMPS employs the well-known power method, taking a random initial state $\ket{\phi_0}$ and applying $\tilde{H}$ iteratively, normalizing at each step to preserve stability, so that

\begin{equation}
\ket{\phi_{n+1}} = \frac{\tilde{H}\ket{\phi_{n}}}{||\tilde{H}\ket{\phi_n}||}. 
\end{equation}
After enough iterations, all that remains will be the dominant eigenvector which we had targeted, $\ket{\psi_{\tilde{\lambda}}}$. Hence, if we could find a way to efficiently apply $\tilde{H}$ to an MPS representation of our initial state, we will have an algorithm that converges to the desired result. Unfortunately, $\tilde{H}$ is generally a complicated, global operator, far from the ``sum of local terms" family of operators that can easily be represented in the matrix product formalism. 

 We do, however, have the operator $O = \tilde{H}^{-1} = H - \lambda$, which \emph{can} still be efficiently represented as an MPO. With this the authors of Ref.~\cite{SIMPS} reimagine the optimization problem of traditional DMRG to perform the desired update. We seek a state $\ket{\phi_n}$ such that (up to normalization), $\ket{\phi_{n+1}} = \tilde{H} \ket{\phi_n}$. In terms of $O$, this state will satisfy

\begin{equation}
O \ket{\phi_{n+1}} - \ket{\phi_n} = 0,
\end{equation}
so we can build an appropriate cost function with an absolute minimum by taking an absolute square. The desired state $\ket{\phi_{n+1}}$ will be the state which minimizes

\begin{equation}\label{eq:SIMPSCost}
\mathcal{L} = ||O \ket{\phi_{n+1}} - \ket{\phi_n}||.
\end{equation}

By recasting this as an optimization problem, it becomes similar to the well-known case of the MPS implementation of Density Matrix Renormalization Group (DMRG) techniques~\cite{DMRGWhite1, DMRGWhite2, DMRGMPS1, DMRGMPS2, DMRGMPSReview}, which can be used to find the ground state of a Hamiltonian by taking the MPS representation of a random initial state and gradually optimizing the set of matrices $\{A^i\}$ at each site. The optimization procedure sweeps back and forth along the chain, until the resulting state has converged. The SIMPS technique proceeds analogously, starting with a random state $\ket{\phi_0}$ and sweeping back and forth to optimize the matrices (specific details of this procedure can be found in Ref~\cite{SIMPS}). Eventually, we converge on a state which satisfies Eq.~\ref{eq:SIMPSCost}. 

When this happens, we have successfully calculated a single application of $\tilde{H}$ to our initial state. This sweeping and optimizing procedure is then repeated over an over, resulting a sequence of states $\{ \ket{\phi_0}, \ket{\phi_1}, \dots \ket{phi_n} \}$, each representing more and more applications of $\tilde{H}$ to the initial state. Eventually, this sequence converges by the power method to our desired, target state: the excited eigenstate of the original Hamiltonian $H$ with energy closest to $\lambda$. The complete algorithm can be thought of as a pair of nested loops, outlined as follows:

\begin{itemize}
\item Initialize a random initial state $\ket{\phi_0}$
\item Loop 1: Iteratively apply $\tilde{H}$ to get $\ket{\phi_{n+1}} = \tilde{H}\ket{\phi_{n}}/||\tilde{H}\ket{\phi_n}||$ until we converge to the target state
\begin{itemize}
 \item Loop 2: In order to compute this application of  $\tilde{H}$, sweep back and forth across the state, optimizing tensors as described above until the resulting state converges
\end{itemize}
\end{itemize}

Because the algorithm depends on two nested loops, we need at least two nested convergence criteria. Our implementation checks multiple quantities for convergence, in order to maximize the efficiency of the algorithm; the full details of our convergence scheme can be found in Appendix ~\ref{AppendixSIMPSConvergence}. But principally, we determine the convergence of the inner loop (in which individual site tensors are optimized to find to find $\ket{\phi_{n+1}} = \tilde{H} \ket{\phi_n}$) by checking the quantity $\delta_1 \equiv |\bra{\phi_n}O\ket{\phi_{n+1}}|$, which clearly should converge to zero when $\ket{\phi_{n+1}}$ is in the desired state because $O = \tilde{H}^{-1}$. As observed in~\cite{SIMPS}, this quantity can be computed essentially for free, because it is equivalent to the overlap $\innerproduct{A_{[j]}}{B_{eff}}$.

To determine whether the outer loop has converged-- that is, to determine whether $\ket{\phi_n} = \tilde{H}^n \ket{\phi_0}$ has converged sufficiently close to the target state, we consider two quantities. The first is simply $\delta_E = |E_n - E_{n+1}|/E_n$, the relative change in the energy of the state (with respect to the \emph{original} Hamiltonian) between steps $n$ and $n+1$. The second, and more important quantity, is the variance of the energy, $\delta_{H^2} = \langle \Delta H^2 \rangle$, which can also be computed efficiently for states represented by an MPS by a variety of methods~\cite{IanBinder1, IanBinder2, WestCumulants}. 
The latter convergence check is particularly valuable since, naturally, we wish to be sure we are studying proper localized eigenstates(rather than simply a superposition of such). Since the energy fluctuation should vanish in a true eigenstate, a strict convergence threshold for this quantity allows us to reject any states whose variance is too large. 

This, then, is the algorithm we will employ when the dimension of our systems becomes to large to compute spectra by exact diagonalization. In both cases (MPS numerics and exact diagonalization) we target a range of states in the middle of the spectrum across many different instances of random disorder. However, it must be noted that, because not \emph{all} quantum states admit a compact and efficient MPS representation (but only states with particular limitations on their correlations such as ground states of local Hamiltonians or MBL eigenstates) we cannot use this procedure to study the eigensates of a Hamiltonian across an broad range of disorders: the eigenstates will only be efficiently computable when we are \emph{within the localized regime}. Hence, we use this numerical data only to probe the transition to localization from above; it does not provide precise evidence of the location of the transition, but rather only helps to bound it from above and to provide additional evidence of localization as we move to the thermodynamic limit.

In our numerical simulations, we employ states with generic, open boundary conditions, and use a bond dimension of 30. We have observed that with a larger bond dimension, such as 50 or 60, fewer applications of $\tilde{H}$ are required before the state converges. However, the increase in the time required to apply each instance of $\tilde{H}$ more than offsets the potential gains. In tests performed for both spin-$1/2$ and spin-1 systems at high disorder, very little difference was seen in either states or observables when comparing states computed at any bond dimension greater than 30, so this smaller value was used to maximize the computational efficiency.
 
 \subsection{Computing Geometric Entanglement} \label{sec:GeomEntMPS}

 In addition to computing the states, we also employ an algorithm based on matrix product states to compute the geometric entanglement of a state. Recall that the geometric entanglement of a state $\ket{\Psi}$ is given as $E_G(\\Psi) = - \log \Lambda$, where $\Lambda$ is the absolute square of the overlap between $\ket{\Psi}$ and the nearest product state $\ket{\Phi}$. Since this definition also involves an optimization problem, we can once again employ a method inspired by DMRG. We start with a random initial state $\ket{\Phi_0}$, but this time explicitly require it to be a \emph{product} state by representing it as an MPS with a bond dimension of just $D = 1$ for all matrices. 

 From this initial state, we iteratively calculate a new state $\ket{\Phi_{n}}$ by sweeping back and forth across the sites of  $\ket{\Phi_{n-1}}$, updating the tensors at each site and seeking to maximize the overlap 
 
 \begin{equation}
\Lambda_{n} =  \frac{|\innerproduct{\Psi, \Phi_n}|}{\sqrt{\langle \Psi | \Psi \rangle \langle \Phi_{n} | \Phi_{n} \rangle }},
\end{equation}
following Ref~\cite{WeiGeometricAlgorithm}. This process is then iterated until the states $\{ \Phi_n \}$ converge. As with DMRG and SIMPS, the optimizing tensors in each step can be computed by solving a generalized eigenvalue problem~\cite{MPSReview}. The resulting overlap between the $n^{th}$ state and the original state is a natural choice for the convergence criterion, since it can be calculated at a minimal computational cost at each step, and is in fact the final quantity of interest. When this sequence of overlaps converges, it need only be squared to produce the quantity $\Lambda$ in Eq.~\ref{eq:GeomEnt}. To avoid becoming numerically trapped at a point which is only locally maximum, we then repeat this procedure for several hundred random initial product states.
 
\begin{figure}\includegraphics[width = 90mm]{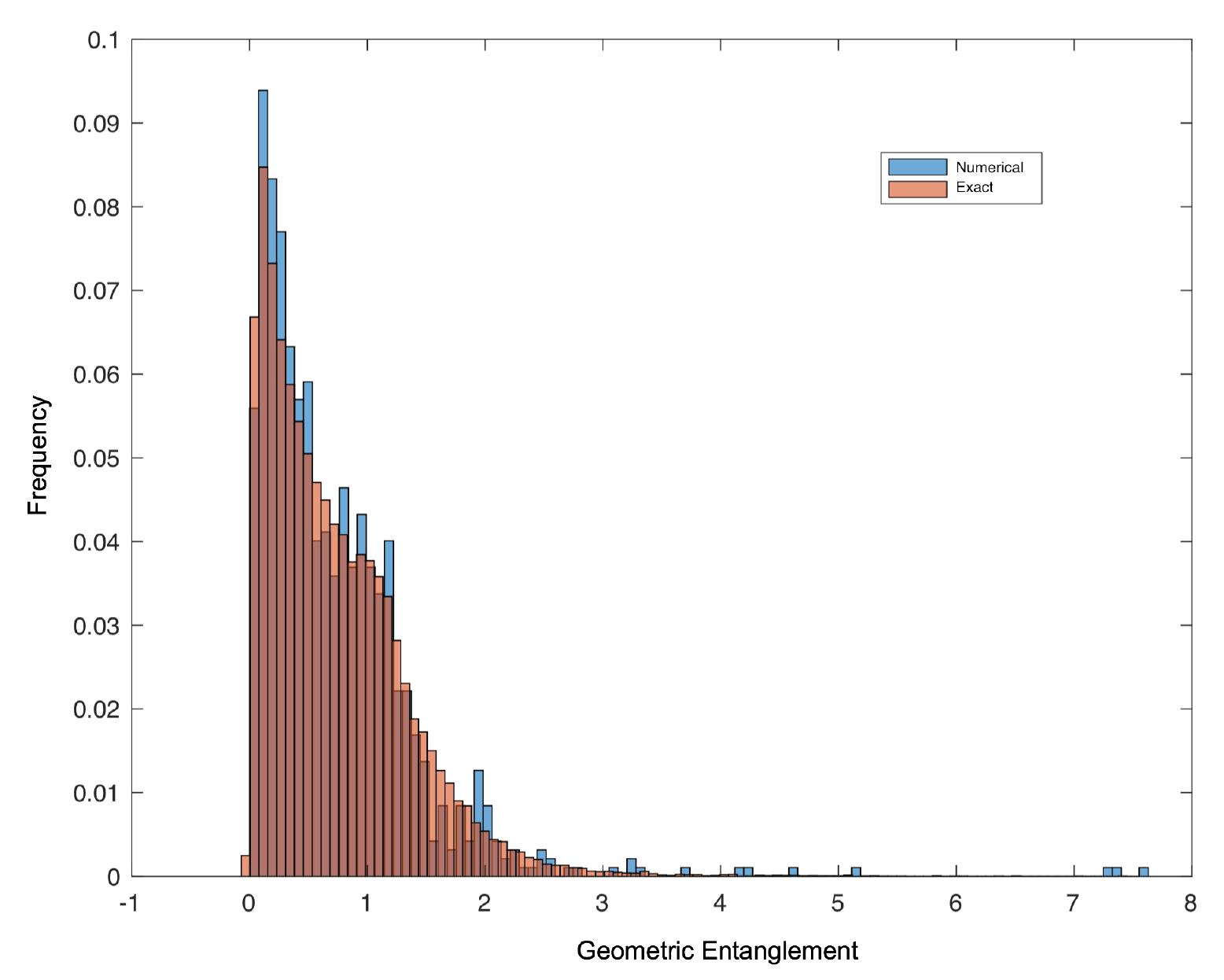}
\caption{\label{fig:GEDistributions}
 (Color online) A histogram of the geometric entanglement of a large collection of numerically generates states ($N \approx 5000$), shown in blue, alongside a similar histogram for the geometric entanglement from states obtained using exact diagonalization, shown in orange. States were obtained from the spectrum of a spin-1/2 disordered Heisenberg system in a strongly disordered region (W = 6). The absence of any systematic difference between the distribution suggests that our numerical procedures do not produce data which are biased towards atypically low entropy states.}
\end{figure}

In Ref~\cite{SIMPS}, it was verified that the matrix product algorithm used to generate these MBL states is not biased with respect to the entanglement as measured by the mid-bond entropy. Specifically, it is shown that the histogram of states generated by the algorithm does not appreciably differ from that found in cases which permit exact diagonalization. This is a vital assurance when studying entanglement properties numerically because numerical algorithms, particularly those based around local updates, can sometimes prefer states with lower entropy. Because of the more global nature of geometric entanglement, and because we compute it here using another optimizing matrix product algorithm involving local updates, it is necessary also to verify that our numerical methods are not biased with respect to geometric entanglement. Specifically, we need to show that if we generate a collection of MBL states using SIMPS, and then compute the geometric entanglement of each using the procedure outlined above, the distribution of geometric entanglement values mirrors that which results from exact diagonalization. This comparison was performed for a variety of spin-1/2 chains of different lengths, and in no instance were the distributions observed to differ in a systematic way. A representative example of these distribution comparisons is given in Fig.~\ref{fig:GEDistributions}. 
 
 \section{MBL and Entanglement in a Spin-1/2 system}\label{SpinHalf}
 
 We begin our exploration of the negativity and geometric entanglement in the same setting as Ref.~\cite{Bera}: a disordered, one-dimensional Heisenberg model on a spin one-half chain, with a Hamiltonian given by 
 
 \begin{equation}\label{eq:RandHeisen}
H_W =  \sum_i  \frac{1}{2} \left( \sigma^x_i \sigma^x_{i+1} + \sigma^y_i \sigma^y_{i+1}+\sigma^z_i \sigma^z_{i+1} \right) + h_i\sigma^z_i,
\end{equation}
where the field strength coefficients $\{h_i\}$ vary from site to site and at each site are drawn randomly from a uniform distribution within the interval $[-W, W]$. The factor of one-half is included to emphasize that this is an instance of an XXZ chain with full-strength interactions in a random field. For this model, which represents perhaps the most widely-studied example of MBL behavior, a considerably body of work (e.g. Refs.~\cite{PalOriginal, Prosen, StdDevMBL, Alet, DisorderCP1, DisorderCP2, Bera}) suggests a transition from ergodic to localized somewhere between $W = 3$ and $W = 4$, with most evidence indicating that it occurs close to $W = 3.7$~\cite{Alet}. Some of the variation in the reported locations of this transition comes from the fact that the tails of the spectrum are likely to localize before the states in the middle~\cite{Huse, Alet}, so that work which looks only at states in the center may report a larger value of the critical disorder strength $W_c$ than work which averages over the entire spectrum. In our work below, we will be looking for localization among states in the middle of the spectrum.

Using a combination of exact diagonalization and MPS methods, we consider the behavior of the total nearest-neighbor negativity (Eq.~\ref{eq:NegNN}) and geometric entanglement (Eq.~\ref{eq:GeomEnt}) for systems of various lengths. For each length, we compute ~50 states distributed around the middle of the spectrum, and average over 1,000 different disordered samples (choices of $\{h_i\}$) for a total of roughly $50,000$ states which are averaged into each data point. We then repeat this process as we sweep across a range of values for the disorder parameter $W$.

\subsection{Negativity}
\begin{figure}
\includegraphics[width = 90mm]{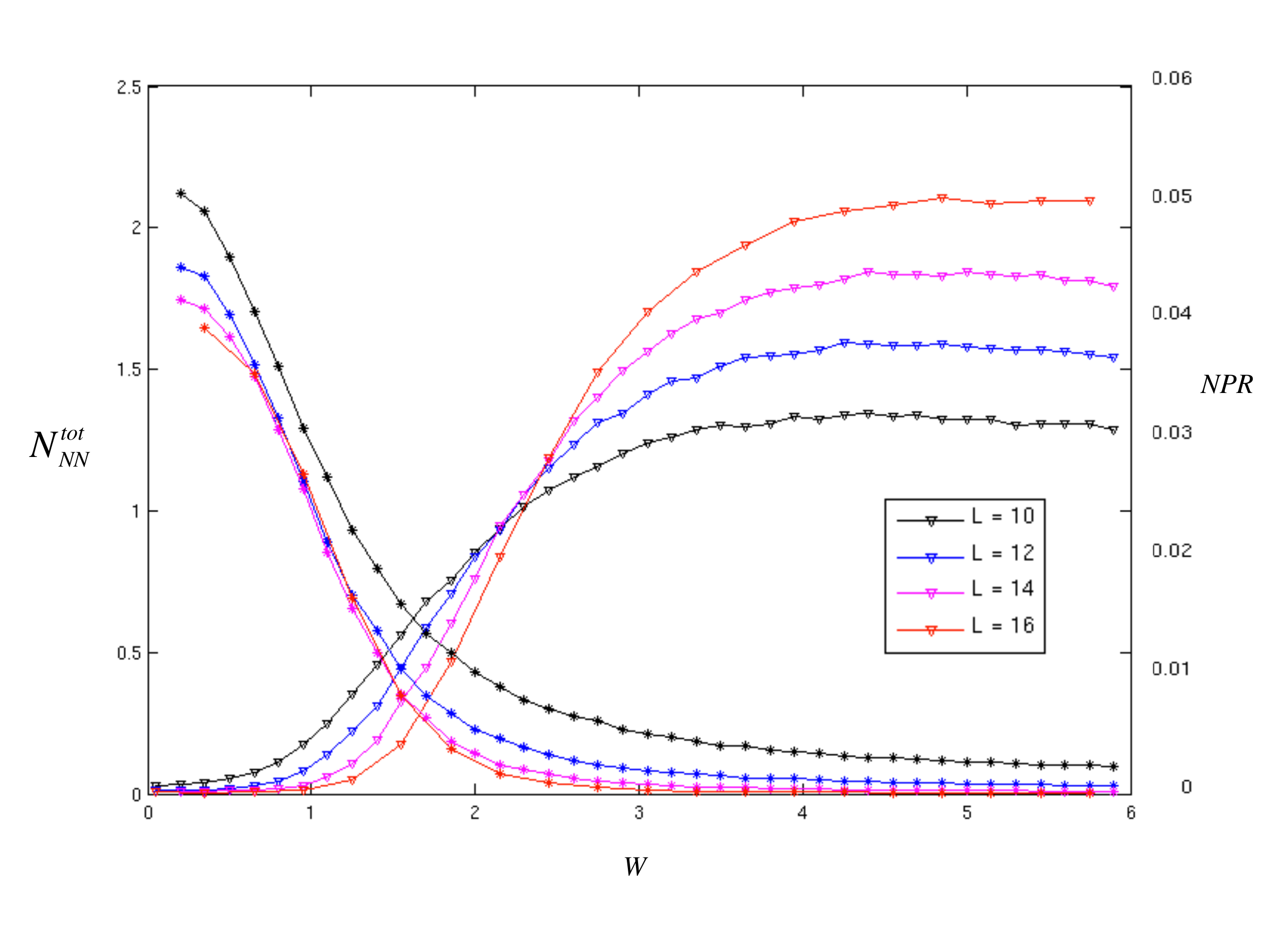}
\caption{\label{fig:Neg} (Color online) The total nearest-neighbor negativity $\mathcal{N}^{tot}_{NN}$(inverted triangles, left axis) and the normalized participation ratios (asterisks; right axis) for the random-field spin-1/2 Heisenberg model, shown for four system sizes. Data are collected using exact diagonalization; 1000 disordered configurations are computed, and for each we compute 50 eigenstates from the middle of the spectrum, where the behavior most resembles the thermodynamic limit. A sharp decrease in the NPR is known to show the transition to the localized phase (estimated for this model to occur around $W_c$ = 3.7~\cite{Alet}), and the negativity is seen to increase significantly just as the NPR decreases. Note that the NPR values here have been normalized by the volume of the Hilbert spaces so that we can compare different system sizes.}
\end{figure}

As shown in Fig.~\ref{fig:Neg}, we find that the total negativity remains small until the anticipated transition, at which point it begins to grow rapidly. As was the case with the concurrence in Ref~\cite{Bera}, the total negativity for points in the localized phase grows with the system size. To show the agreement of this behavior with other known measures of localization, we have also plotted the Normalized Participation Ratios (NPR), averaged over disorder. The NPR (and related quantities such as participation entropy) are widely used as reliable indicators of localization~\cite{ParticipationEntropyOriginal, ParticipationEntropy2, Alet}. For a general state $\ket{\psi}$ expanded in some configuration basis $\ket{i}$, so that $\ket{\psi} = \sum_i c_i \ket{i}$, the NPR is defined as

\begin{equation}
P(\psi) = \frac{\left( \sum_i |c_i|^2 \right)^2 }{\sum_i |c_i|^4}. 
\end{equation}
Observe that for a normalized vector, this is simply $1 / \sum_i p_i^2$, where $p_i$ is the probability for $\ket{\psi}$ to be in the configuration $\ket{i}$. Hence the NPR ranges from $P(\psi) = 1$ when the system is completely localized to a single state, to $P(\psi) = N$ when the state is uniformly distributed across all $N$ possible configurations with probability $p_i = 1/N$ for each. Note that for instances where the volume of the relevant Hilbert space is not fixed, one may wish to normalize the NPR by including this factor before comparing the NPR of different states~\cite{QPMBL}.

It was also observed in Ref.~\cite{Bera} that there is an additional indicator of localization which can be found in details of the entanglement. In the thermalized regime where the ETH is satisfied, the concurrence between two spins is negligible and essentially independent of the distance between them. But in a localized system, the bipartite entanglement between two sites (as averaged over disorder) should be suppressed exponentially by the distance between the two sites. This behavior is uniquely characteristic of the localized regime and hence forms a second kind of entanglement-based indicator of MBL. Furthermore, it should be possible to extract a characteristic entanglement length $\xi$ from the scaling relationship, implicitly defined by the relation
\begin{equation}
\mathcal{C}(\rho_{i,i+d}) \propto \mathcal{C}(\rho_{i,i+1}) e^{-d/\xi_\mathcal{C}}
\end{equation}
for concurrence or

\begin{equation}
\mathcal{N}(\rho_{i,i+d}) \propto \mathcal{N}(\rho_{i,i+1}) e^{-d/\xi_\mathcal{N}}
\end{equation} 
for negativity.

\begin{figure}
\includegraphics[width = 90mm]{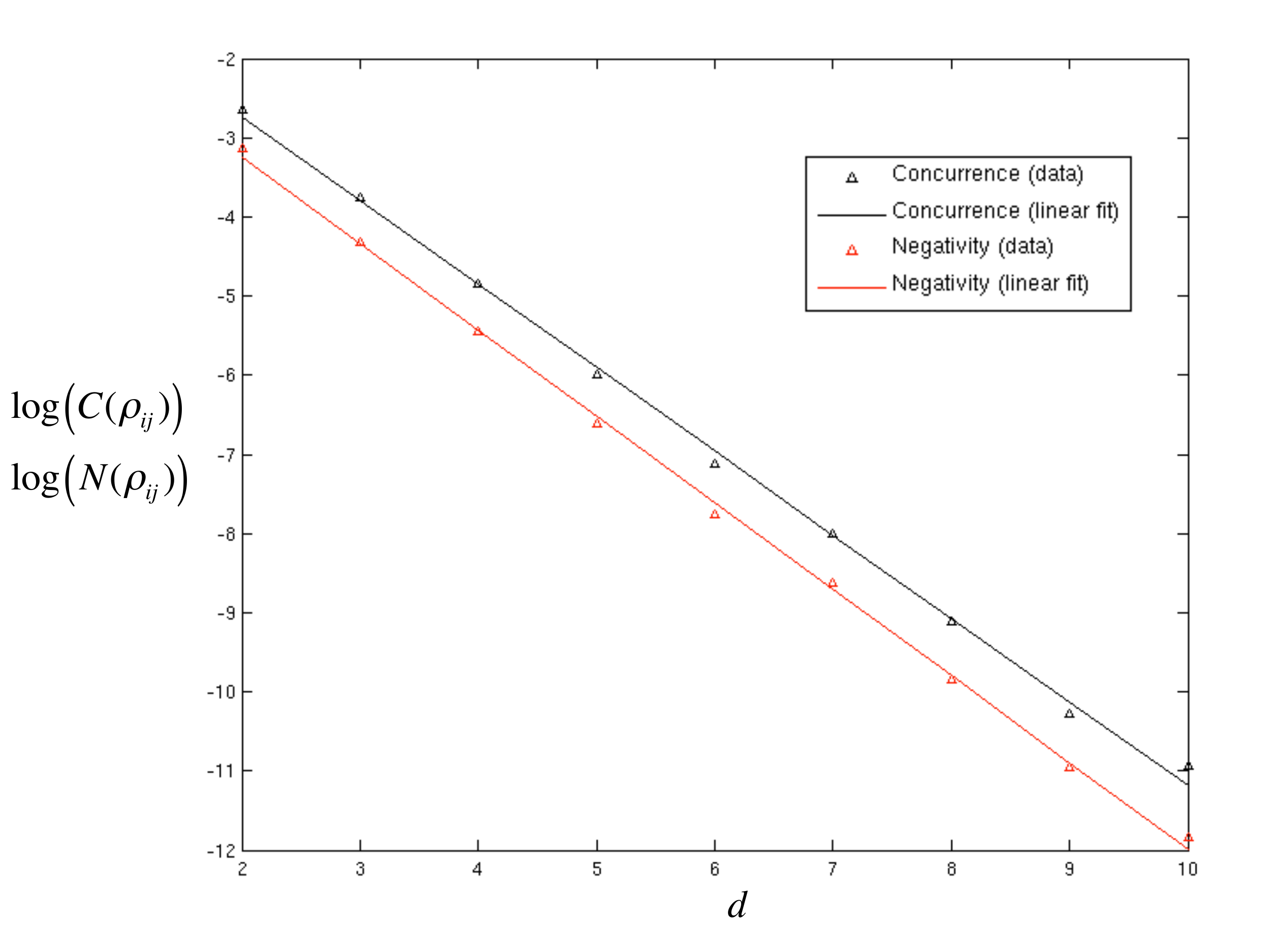}
\caption{\label{fig:CCRNegDecay} (Color online) Using Data from SIMPS, we compute the concurrence and negativity between sites $i$ and $j$ in a disordered Heisenberg chain of length $20$ in the localized regime ($W = 6$). Both quantities decay exponentially with the distance $d = |i-j|$ between the sites; note that here we have plotted the log of each (concurrence in black, negativity in red). Fit lines are included as a guide to the eye and to demonstrate that the rate of decay is very similar for both negativity and concurrence, suggesting that both quantities capture the same characteristic entanglement length. The slope of the concurrence fit line is $\xi_{\mathcal{C}} = -1.08$; for negativity it is $\xi_{\mathcal{N}} = -1.12$. This behavior is representative of other system lengths as well, although the slopes appear to have some weak dependence on the strength of disorder. Note that we show here only distances out to $d = L/2$; beyond this point the entanglements are so close to zero that they are very sensitive to noise and finite size effects, as discussed in Ref.~\cite{Bera}.  }
\end{figure}

For the systems we have considered, it appears that the negativity shares not only a similar exponential form, but may have the same entanglement length, based on our result that  $\xi_{\mathcal{C}} = -1.08 \pm 0.034 and \xi_{\mathcal{N}} = -1.12 \pm 0.05$. In Fig.~\ref{fig:CCRNegDecay}, for example, we show (in semilog scale) the concurrence and entanglement (averaged over disorder) for a system of length $L = 20$, with states computed from SIMPS. Up to the level of the statistical noise resulting from the disordered average, the lines have very comparable slope. This is representative of similar behavior seen for shorter systems as well. This again suggests that the negativity can also be used in the same manner as the concurrence as an indicator of localization. 

A In Ref~\cite{Bera}, scaling relationship was also found empirically between the length of the system second derivative of the average concurrence, $d^2\mathcal{C}^{avg}_{NN}/dW^2$ , with the scaling collapse used to extract a specific estimation of the localized-delocalized transition. Note here that we are using now the per-site average of the concurrence rather than the total concurrence. In particular, it was shown that there is a universal scaling function $\Phi$ for which data collapse can be observed from, namely

\begin{equation}\label{eq:MBLCollapse}
\frac{1}{L^a}\frac{d^2\mathcal{C}^{avg}_{NN}}{dW^2} = \Phi \left( L^b (W - W_c) \right),
\end{equation}
with parameters $a \approx 0.5$, $b \approx 0.6$, and $W_c = 3.7$. This last number is in good agreement with the general consensus on the location of the transition for this model~\cite{Alet}.

Naturally, we wish to see if the same holds true for the average negativity, $\mathcal{N}^{avg}_{NN}$. Our negativity data points are all averages over disorder and are hence already somewhat noisy, and taking a numerical derivative of a noisy curve will amplify the noise significantly. Therefore, to reduce the level of statistics required to see the relationship, we can instead perform a numerical fit of our data, and consider the derivatives of the resulting polynomial. 

\begin{figure}
\includegraphics[width = 90mm]{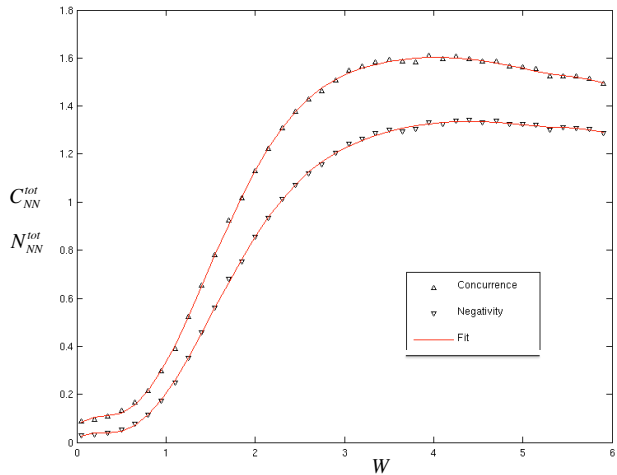}
\caption{\label{fig:CCRNegFits}  (Color online) L = 10 data for concurrence (triangles) and negativity (inverted triangles) are presented together, along with numerical fits from a polynomial of degree 9. The fit is in good agreement with the qualitative behavior of the data, particularly in the crucial transition region.}
\end{figure}

To avoid overfitting, we have optimized the degree of the fitting polynomial by considering the ratio

\begin{equation}
\frac{\sum_i^n r_i^2}{(n-m-1)},
\end{equation}
where the $r_i$ are the residuals after the fit, $n$ is the number of data points, and $m$ is the order of the polynomial. We choose $m$ so that this ratio is at a minimum or appears to have converged; we also compare the resulting for various subsets of the data to ensure it remains relatively stable. The data in the relevant domain can be fit very well by a polynomial of degree 9; see for example Fig.~\ref{fig:CCRNegFits}. Note that we do not need to extrapolate outside the fit domain (and because of potential issues interpolating and taking derivatives at the endpoints of the original data, we often omit these points in our finite size scaling studies).

\begin{figure}
\includegraphics[width = 90mm]{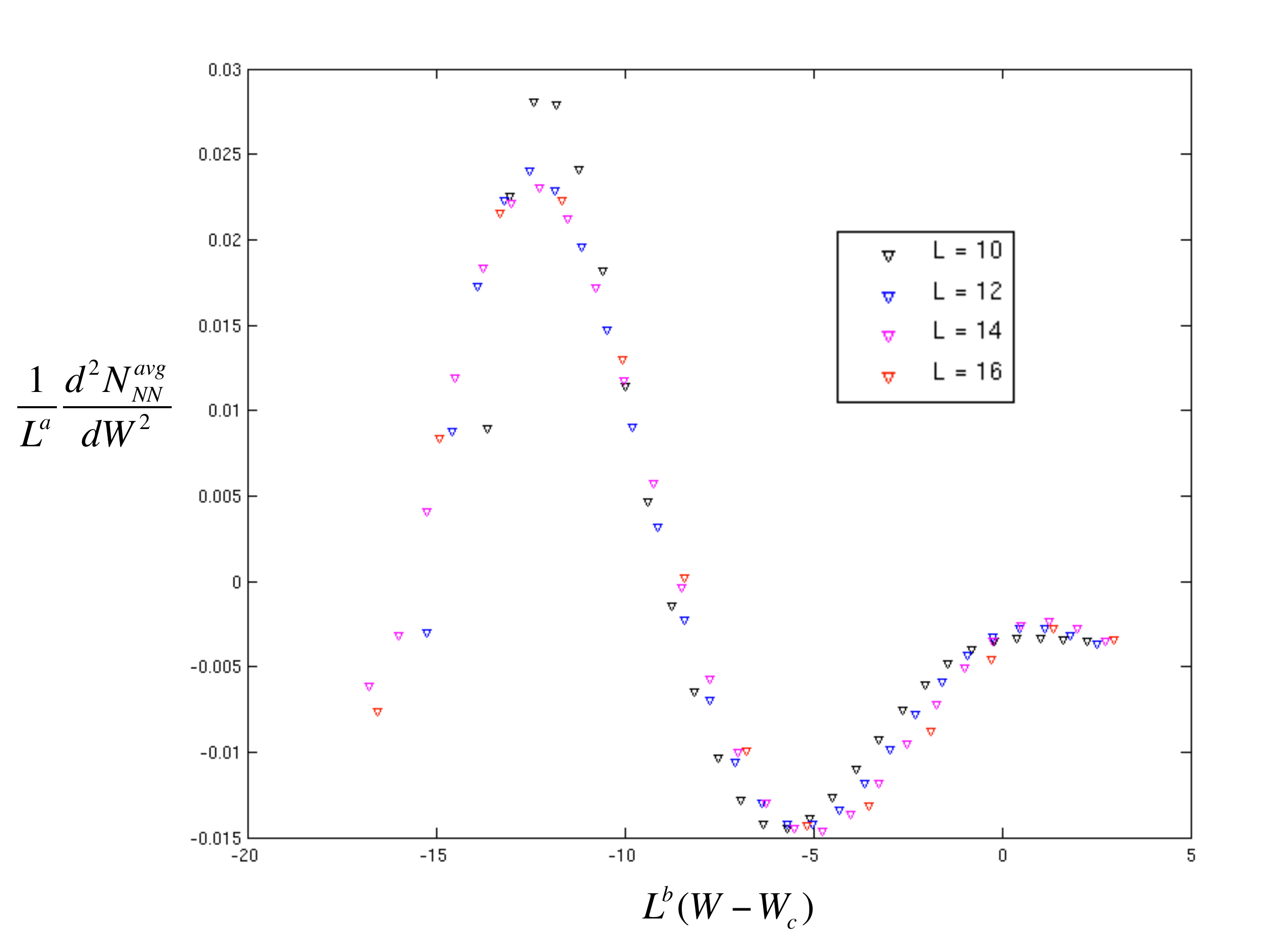}
\caption{\label{fig:NegCollapse}  (Color online) Total nearest-neighbor negativity data is shown for various system sizes in relation to the scaling function in Eq.~\ref{eq:MBLCollapse}. Comparing the result here to the results in Ref.~\cite{Bera}, we see that both the concurrence and the negativity show very similar scaling behavior, suggesting that both can be used to detect the localized to delocalized transition point.}
\end{figure}

Using these numerical derivatives and the same scaling parameters as above, we can reproduce the evidence of universal scaling seen in the concurrence, suggesting that the fitting prodecude is sufficiently accurate to provide evidence about the MBL transition. More importantly, the same scaling parameters also show a scaling relationship in the negativity, as shown in Fig.~\ref{fig:NegCollapse}. The data collapse is particularly evident for systems with $L > 10$, likely because the smallest system is the most subject to finite-size effects. Since this collapse disappears as we move away from these particular values of $a, b$, and $W_c$, it appears that this technique offers yet another way to either locate or verify the location of the localized-delocalized transition, and shows that it can be done with negativity as well as with concurrence.
 
\subsection{Relationship between Negativity and Concurrence}

\begin{figure}
\includegraphics[width = 90mm]{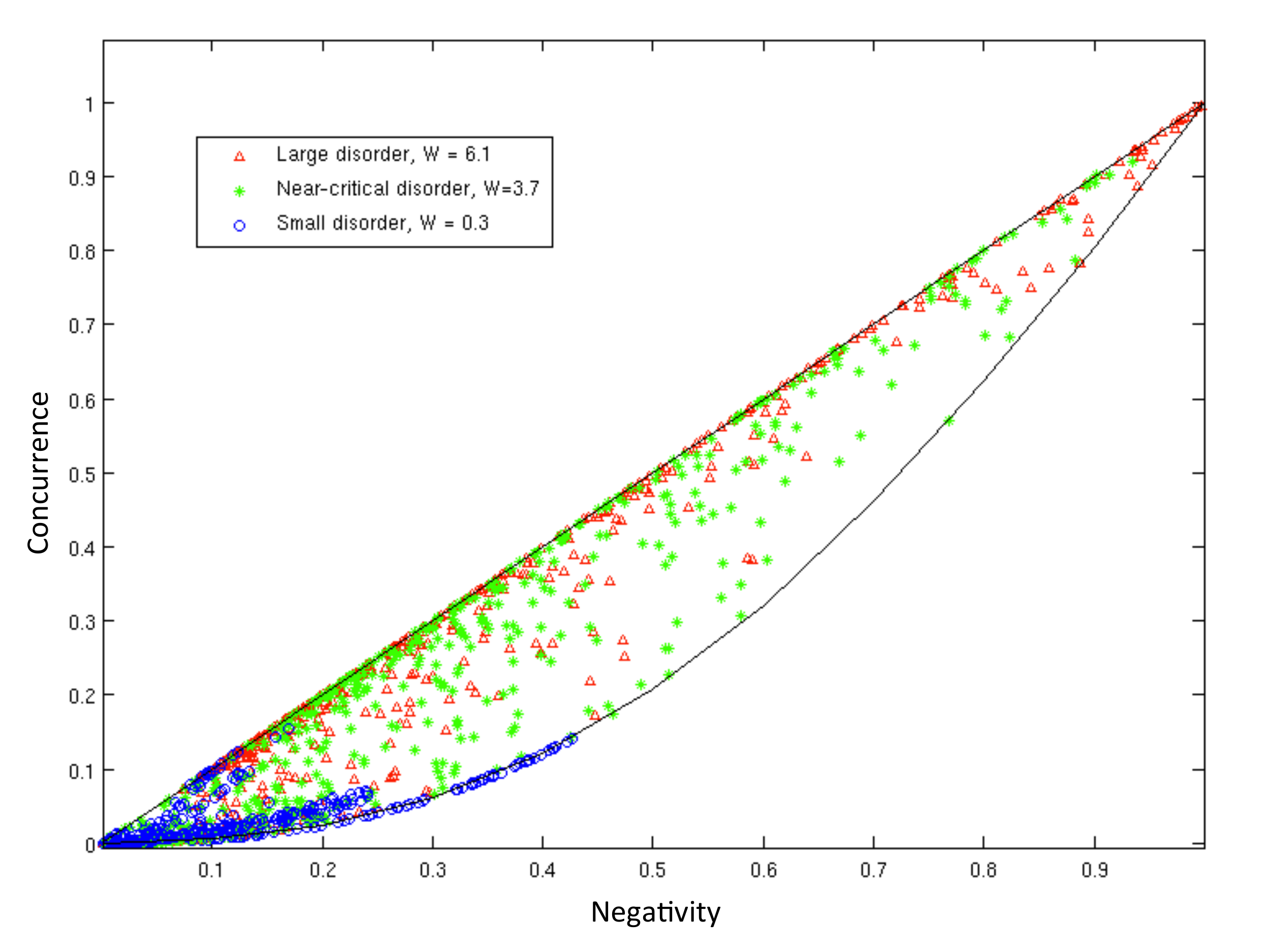}
\caption{\label{fig:NegBounds} (Color online) Negativity and concurrence are compared for different disordered samples in an $L = 12$ system (data from exact diagonalization). Black lines indicate the analytic bounds on the relationship between concurrence and negativity for the two-quibit case; the negativity is never larger than the concurrence, and always larger than $\sqrt{(1-\mathcal{C})^2+\mathcal{C}^2} - (1-\mathcal{C})$. Qualitatively, we see that in subsystems from delocalized states (blue circles), both quantities are relatively small, and the lower bound is more likely to be saturated. In the case of strong disorder (red triangles) the full range of possible values for each is explored, with the upper bound more likely to be saturated. The intermediate case near the localized/delocalized transition shows behavior much more akin to the fully localized case, though this may be the result of finite size effects which tends to increase localized behavior in small systems.}
\end{figure}

\begin{figure}
\includegraphics[width = 90mm]{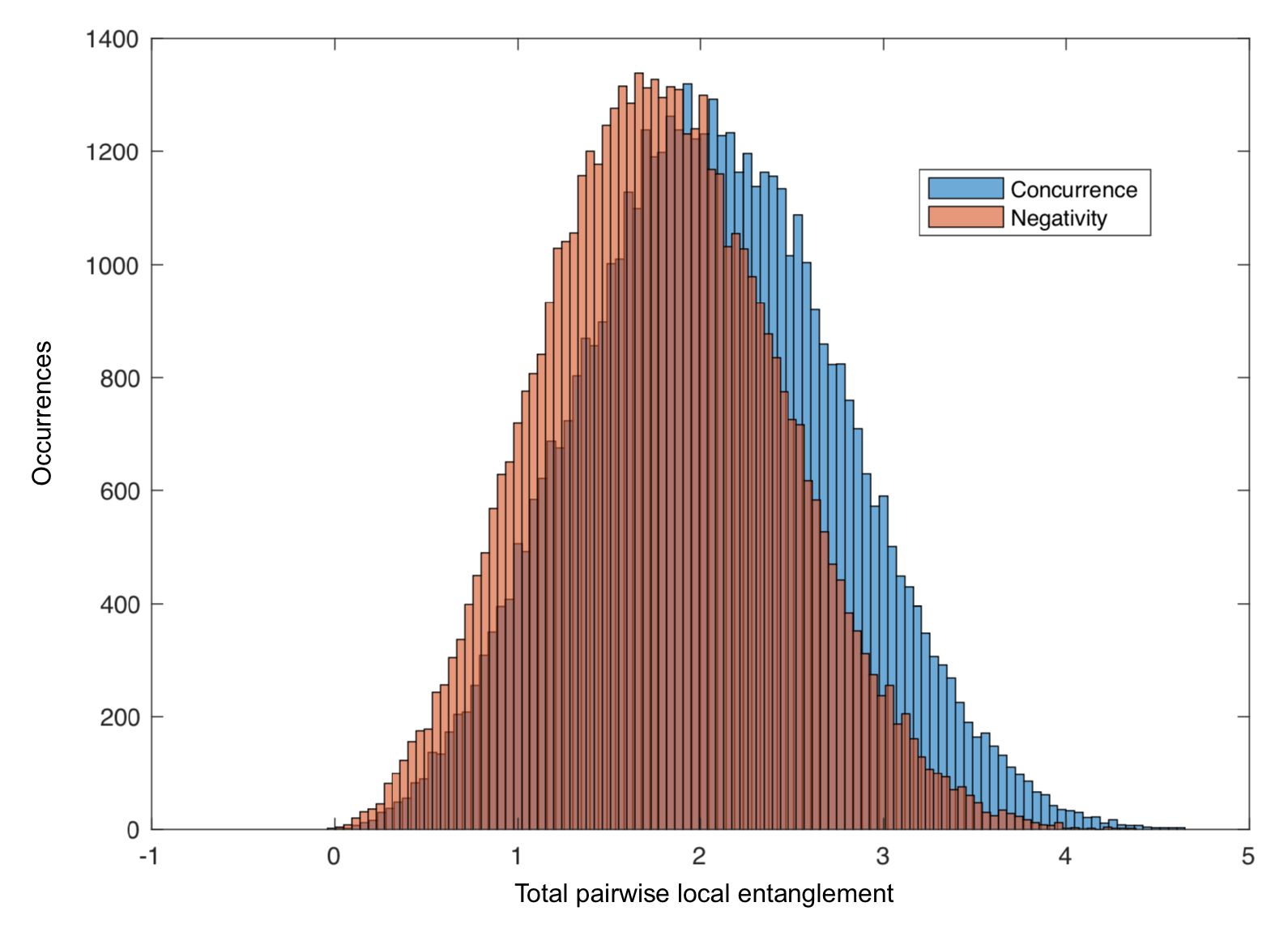}
\caption{\label{fig:CCRvsNEGDistributions} (Color online) Distributions of total negativity and concurrence compared across the eigenstates of an $L = 16$ system with strong disorder (W = 6). Both distributions are qualitatively quite similar and appear to be normally distributed, with the concurrence distribution centered slightly higher, as one might expect given that the negativity of a qubit-qubit state never exceeds its concurrence. }
\end{figure}

Two comments must briefly be made on the relationship between the negativity and concurrence. For the case of two-qubit systems, strict analytical bounds exist for the relationship between these two measures. While both quantities cover the interval $[0, 1]$, the negativity is always bounded above by the concurrence, and bounded below by~\cite{NegCCRComparison}

\begin{equation}
\mathcal{C} > \mathcal{N} > \sqrt{(1-\mathcal{C})^2+\mathcal{C}^2} - (1-\mathcal{C}).
\end{equation}
From these relations, it is clear why the negativity also displays exponential decay with distance in the localized phase, since it is bounded above by an exponentially decaying quantity, although it is not immediately obvious that the decay constants should be the same. The existence of the lower bound in this case also helps to motivate the similarity in the qualitative behavior across the transition. We note (see Fig.~\ref{fig:NegBounds}) that states in the delocalized regime seem more likely to saturate the lower bound, whereas localized states are more likely to saturate the upper bound where the two quantities are equal.  
The relationship between these two quantities can also be probed by considering the distributions of each across the eigenstates of some particular disordered chain. One such comparison, for the particular instance of a $W = 6$ disordered system of length $12$, is shown in Fig~\ref{fig:CCRvsNEGDistributions}. The shape of the distributions, and the relationship between the two, is illustrative of that which we have also seen for other values of $W$ above $W_c = 3.7$. From the cases we have studied, it appears also that the difference between the central values of the distributions grows smaller as the disorder increases, consistent with the above observation that for greater disorder, it becomes more common for eigenstates to display equality in the two entanglement measures. However, as we have examined this only for a specific system, a more rigorous investigation of this relationship may be a topic for future study.

\begin{figure}
\includegraphics[width = 90mm]{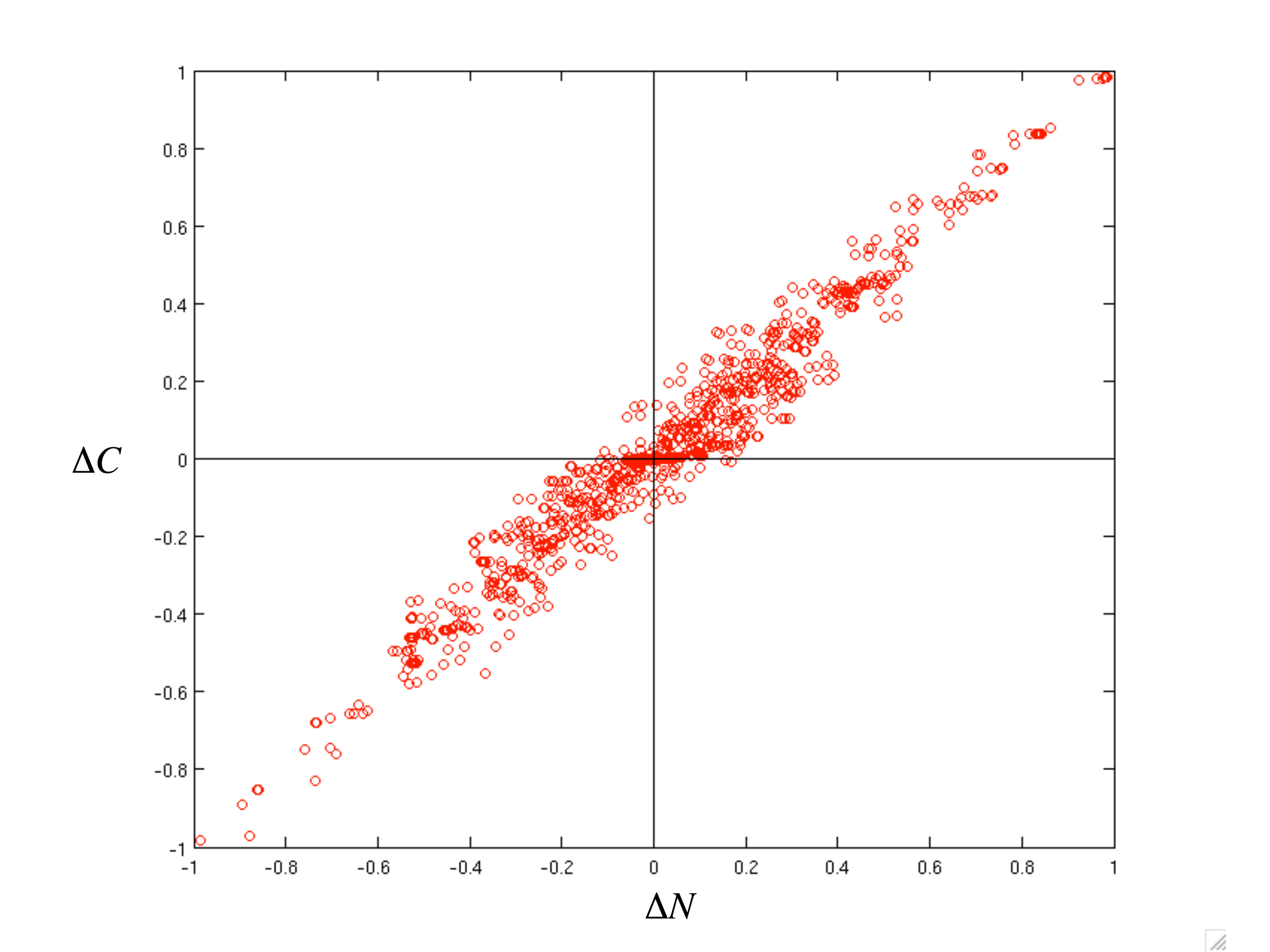}
\caption{\label{fig:Scatter} (Color online) The ordering of concurrence and negativity is considered by comparing equivalent two-qubit subystems from pairs of different states in the disordered ensemble. The difference between concurrences is plotted versus the difference in negativities. Thus, data points in quadrants $I$ and $III$ represent ``well-ordered" pairs, in which the state with the larger concurrence also has the larger negativity. The presence of data points in quadrants $II$ and $IV$ shows also that ``ill-ordered" subsystems are possible, though not common. Data plotted here are from a system of  $L$ = 20 and $W = 6$, though the pattern is typical for states, at least in the localized regime. }
\end{figure}

The other important fact about the relationship between concurrence and negativity is that the two measures, while closely related, do not share the same ordering of entanglement. In other words, consider a pair of two-qubit states $\rho_1$ and $\rho_2$; if we observe that $\mathcal{C}(\rho_1) > \mathcal{C}(\rho_2)$, it does \emph{not} necessarily follow that $\mathcal{N}(\rho_1) > \mathcal{N}(\rho_2)$ ~\cite{NegCCROrdering1, NegCCROrdering2}. We have found that, among the two-qubit subsystems of states in the disorderd Heisenberg model, such ``ill-ordering" are uncommon, but hardly rare (see Fig.~\ref{fig:Scatter}). The fraction of the subsystems showing this property seems relatively constant with disorder strength; this area is left as a direction for future study.

\subsection{Geometric Entanglement}
 
\begin{figure}
\includegraphics[width = 90mm]{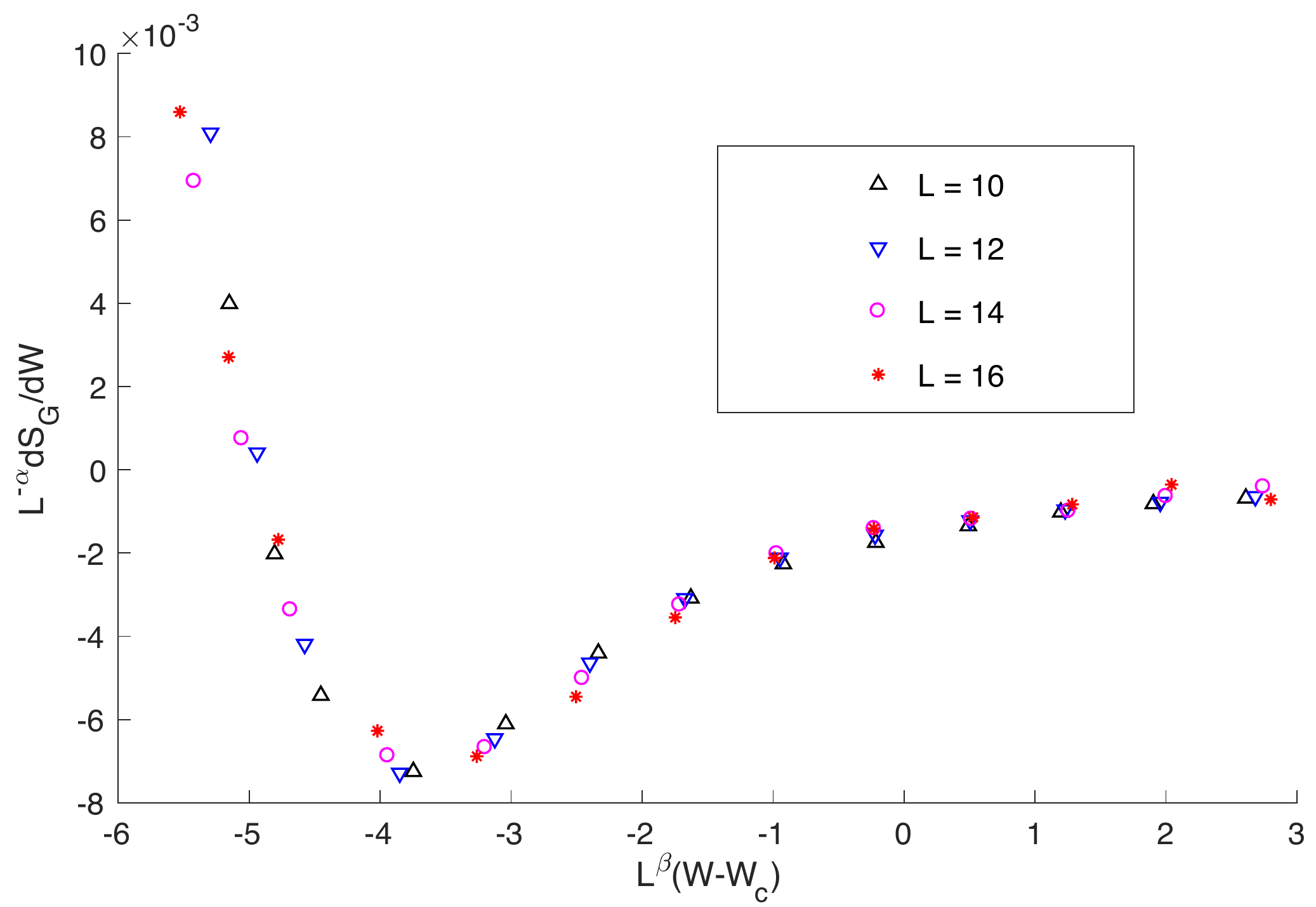}
\caption{\label{fig:SpinHalfGECollapse} (Color online)  The geometric entanglement data from the random-field spin-1/2 Heisenberg model, plotted for various system sizes in relation to the scaling function defined by Eq.~\ref{eq:GECollapse}. Although the nature of the scaling relationship is different than for the cases of concurrence and negativity, its existence nevertheless provides additional evidence for the location of a transition around $W_c = 3.7$, and demonstrates that the geometric entanglement, too, can be a valuable tool for studying localized behavior in spin chains.}
\end{figure}

 Unlike the negativity and concurrence methods described above, our approach to the geometric entanglement provides measure of global entanglement. Unsurprisingly then, its behavior is found to be qualitatively opposite of the other two entanglement measures we have considered. As shown in Fig.~\ref{fig:SpinHalfGECollapse}, the geometric entanglement of the spin-1/2 system (averaged over disorder) is quite large when the disorder parameter is small, but decreases as the disorder is turned up, and more of the entanglement becomes concentrated into localized pockets. In this manner it is also reminiscent of the NPR. 
 
\begin{figure}
\includegraphics[width = 90mm]{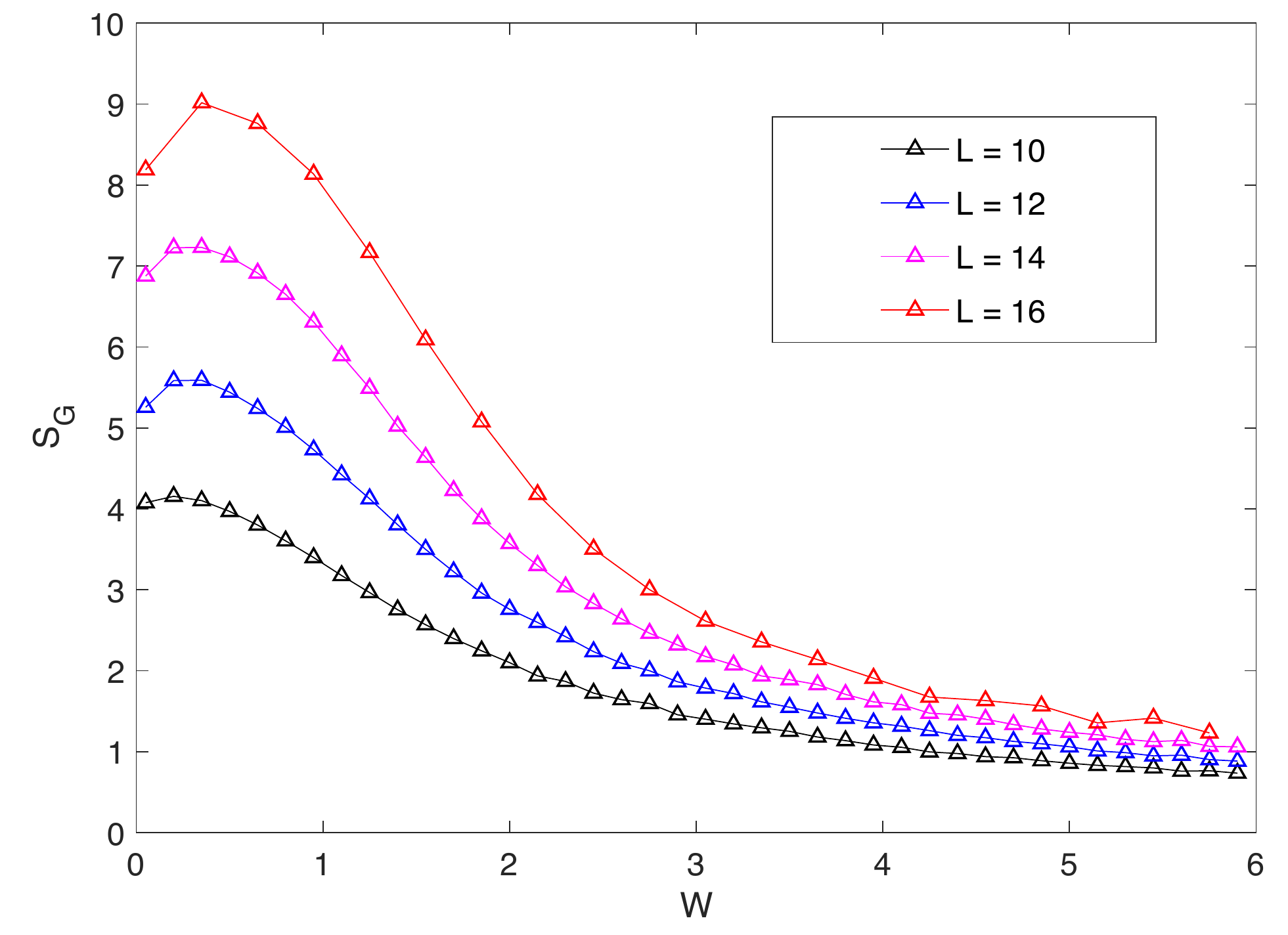}
\caption{\label{fig:SpinHalfGE} (Color online). The geometric entanglement (averaged over disorder) of states, for the random-field spin-1/2 Heisenberg model, shown for four system sizes. Data are collected using exact diagonalization and the MPS method for computing geometric entanglement described above. We compute 1000 disordered configurations, and for each we compute 50 eigenstates from the middle of the spectrum, where the behavior most resembles the thermodynamic limit. As expected, the global entanglement decreases as the disorder increases, and entanglement becomes more concentrated in localized pockets. }
\end{figure}
 
As with the concurrence and negativity, the change in the geometric entanglement is fairly gradual, and hence on its own is not a good candidate for identifying the \emph{location} of the transition. However, we can still consider the finite size scaling law in the same manner as above. In this case, even searching over a large parameter space in $a$, $b$, and $W_c$, we have not observed a clear data collapse of the form given in Eq.~\ref{eq:MBLCollapse}, although we do not mean to imply that we have definitively excluded such scaling. However, we have observed a comparable scaling relationship in terms of the \emph{first} derivative of the geometric entanglement. Since the first-order derivative may serve as an indicator of phase transitions (as seen also in other contexts~\cite{GEFirstDerivative}), unlike the second-order derivatiev needed in the case of concurrence and negativity, this is potentially an advantage to the use of geometric entanglement as an indicator.

The specific form of the apparent scaling relationship is given by

\begin{equation}\label{eq:GECollapse}
\frac{1}{L^\alpha}\frac{d S_G}{dW} = \Phi \left( L^\beta (W - W_c) \right).
\end{equation}
 
For the parameters $\alpha = 2 \pm 0.1$, $\beta = 0.15 \pm 0.1$ and $W_c = 3.7 \pm 0.1$, we observe a clear, consistent data collapse on both sides of the transition, as shown in Fig.~\ref{fig:SpinHalfGECollapse}. Hence, the behavior of the geometric entanglement is also consistent with that of other localization indicators when it comes to identifying the location of the transition.
 
%\begin{figure}
%\includegraphics[width = 90mm]{SpinHalfGE.pdf}
%\caption{\label{fig:SpinHalfDGE} (Color online)  }
%\end{figure}

 \section{MBL and Entanglement in a spin-1 system}\label{Spin1}
 The reason for using negativity and geometric entanglement in addition to concurrence is the ability to compute them in a straightforward manner even for systems with higher local dimension. Therefore, we wish also to demonstrate the application of these techniques to a spin chain model with higher spin. We choose the spin-1 Heisenberg model in a random field; an immediate analogue to Eq.~\ref{eq:RandHeisen}, given by 
 
  \begin{equation}\label{eq:Spin1Heisen}
H_W =  \sum_i  \frac{1}{2} \left( S^x_i S^x_{i+1} + S^y_i S^y_{i+1}+S^z_i S^z_{i+1} \right) + h_i S^z_i,
\end{equation}
where $S_{\{x, y, z\}}^{i}$ are the spin-1 matrices acting at site "i", and as before, the random field strengths $h_i$ are drawn from a uniform distribution between zero and $W$ Note here that the factor of $1/2$ has been included simply for convenience in comparing to the spin-$1/2$; without it, the results given here for various values of $W$ simply double. 

 As remarked above, the increases in memory demands and in the scaling of the Lanczos method mean that a spin-one model such as this becomes much more computationally expensive to solve exactly. To study this model, we will use a combination of exact diagonalization for short chains, and SIMPS for longer chains, to reach up to size 14, well beyond what we could achieve with exact diagonalization alone. However, because the SIMPS algorithm can only be expected to converge a state in a \emph{localized} phase, for the longer chains we are unable to sweep across the entire range of disordered values. Instead, we must start at large disorder and attempt to sweep backwards as far as possible. This limitation, combined with stronger finite size effects for the slightly shorter systems, make it more difficult to give precise estimates of the transition. However, our results still give a clear indication that a localized regime exists, and can still provide guidance as to the value of critical disorder $W_c$.
 
In the data given below, approximately ten thousand states are used in the disorder averages for the shorter chains, and approximately five thousand for the longer chains (the number varies slightly because a small number of states which failed to converge to an eigenstate based on the $\Delta_{H^2}$ criterion were rejected from each dataset).

\begin{figure}
\includegraphics[width = 90mm]{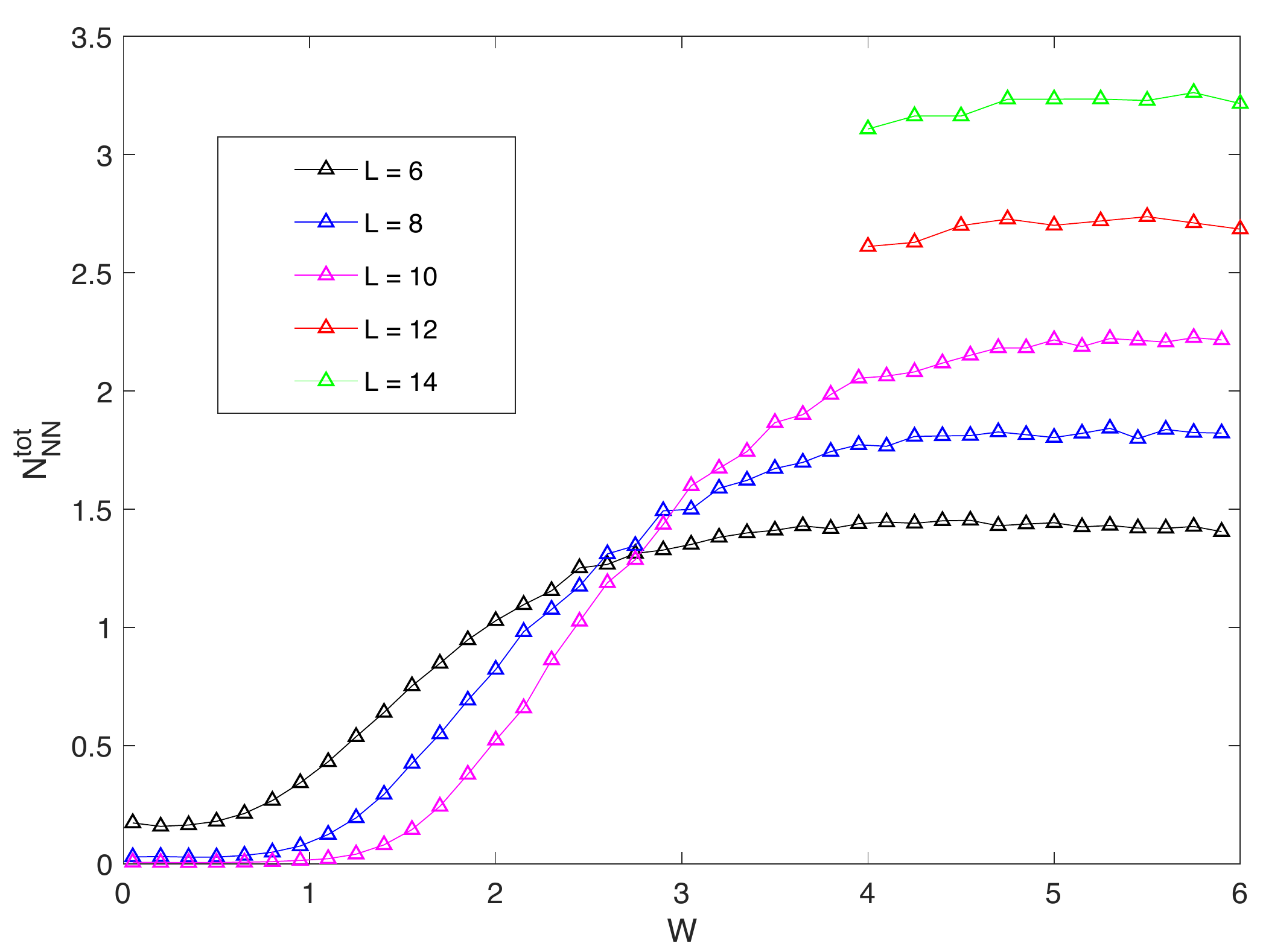}
\caption{\label{fig:Spin1NegNN} (Color online)  The total nearest-neighbor negativity $\mathcal{N}^{tot}_{NN}$ for the random-field spin-1 Heisenberg model, shown for six system sizes. Data are collected using a combination of exact diagonalization and SIMPS, with ~5,000-10,000 states used in each system size (generally fewer states at the larger sizes). States are computed from the middle of the spectrum, where the behavior most resembles the thermodynamic limit. The pattern is very similar to the one seen in the spin-1/2 case, with a sudden increase in local entanglement as the disorder increases, reaching a plateau that scales with the system size.}
\end{figure}
 
 The first evidence that a localized regime exists can be seen clearly in Fig.~\ref{fig:Spin1NegNN}, as the total nearest-neighbor negativity is again seen to be initially near zero, with a sharp increase as the disorder strength increases, eventually reaching a plateau which scales with the system size. It should be noted that the negativity between two spin-1 particles may not completely capture any entanglement present; this is because the Peres-Horodecki criterion upon which is it based is necessary, but not sufficient, to establish separability of the density matrix for the $3 \times 3$ case~\cite{Horodecki}. Hence, it is possible there is \emph{additional} entanglement between the nearest neighbors not captured by the negativity in Fig.~\ref{fig:Spin1NegNN}; but this would not change the clear pattern of entanglement which both increases with disorder and with the system size. The evidence suggests that the system displays localized behavior for roughly $W_c \geq 4$. This number is based primarily on the observation that it becomes numerically difficult to produce a converged MPS state (with a success rate of less than 15\%) for values of W smaller than 4. Since it is plausible that our algorithm is able to converge some states below the transition, albeit with lesser precision, this can be interpreted as a lower bound on $W_c$ for this model.
 
\begin{figure}
\includegraphics[width = 90mm]{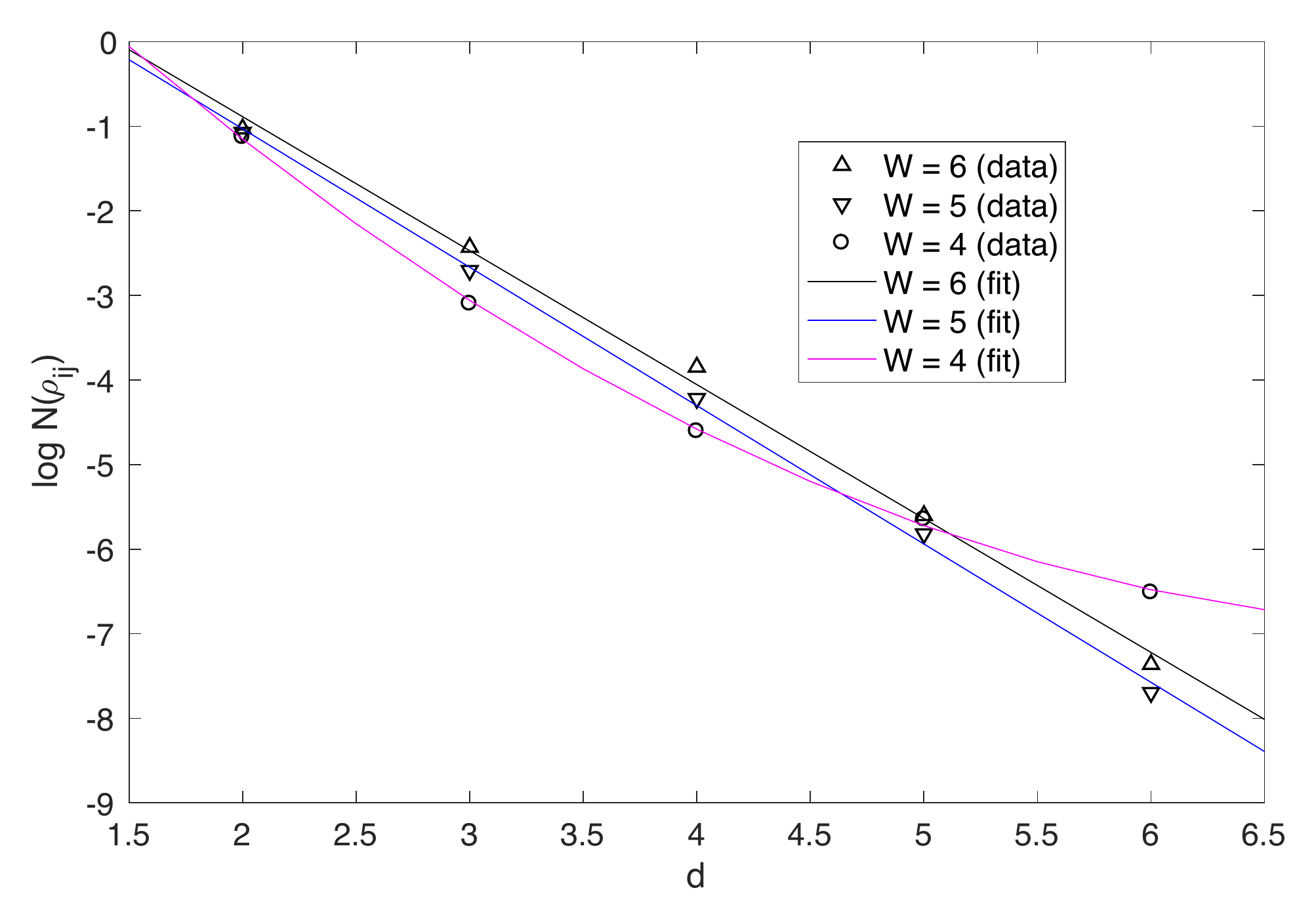}
\caption{\label{fig:Spin1NegDecay} (Color online) Using Data from SIMPS, we compute the average negativity between sites $i$ and $j$ in a disordered, spin-1 Heisenberg chain of length $14$ for several values of the disorder strength parameter $W$. When the disorder is large, the entanglement can be seen to fall exponentially with the distance $d = |i-j|$ between the sites. When the disorder becomes smaller, around $W = 4$, the decay becomes lessened, suggesting that $W_c \in (4, 5)$. For very small values of disorder (not plotted), there is no relationship between the entanglement and the system length because the entanglement is negligible in all cases. As above, we show here only distances out to $d = L/2$.}
\end{figure}

Further evidence for the existence of localized and delocalized phases can be found in the behavior of the negativity between spins which are not nearest neighbors. As we did for the spin-1/2 model, we can plot the decay of the entanglement between sites $i$ and $j$ as a function of the distance $d_{ij}$ between them. Even for the large system sizes, there is a clear exponential decay in the entanglement when the disorder is sufficiently large, with an entanglement length of approximately $\xi_E = 0.6$. On the other hand, as the disorder becomes weaker, the decay begins to take a sub-exponential form. Ultimately, for very low disorder in the short chains, we see uniformly negligible entanglement as in the spin-1/2 case, consistent with a delocalized phase. In Fig.~\ref{fig:Spin1NegDecay}, we examine this behavior for $W$ ranging between $4$ and $6$. The slightly sub-exponential nature of the decay at $W = 4$, compared with the more clearly exponential behavior at $W = 5$, suggests a transition point somewhere between the two, $W_c \in (4, 5)$.

 To attempt a more rigorous identification of the transition point, we can also consider the finite-size scaling of this negativity. Because of the qualitative similarity in the behavior of the negativity for the spin-1/2 model, we hypothesize the same scaling behavior as described above (see Eq.~\ref{eq:MBLCollapse}; we have used here ``a" and ``b" instead of ``$\alpha$" and ``$\beta$" to distinguish the spin-1/2 case). Sweeping over a portion of parameter space, including $W_c \in (4, 5)$, we find the best data collapse for $W_c = 4.7$, $a = -1.1$, and $b = 0.5$. We stress that all of these numbers, but particularly $W_c$, are approximate, as collapse is difficult to identify quantitatively, and because small changes to these values do not substantially change the result. However, the results do suggest a slightly larger value of  $W_c$ than can be identified from the raw negativity data alone, and are most consistent with $W_c \in (4.5, 5)$.
 
  \begin{figure}
\includegraphics[width = 90mm]{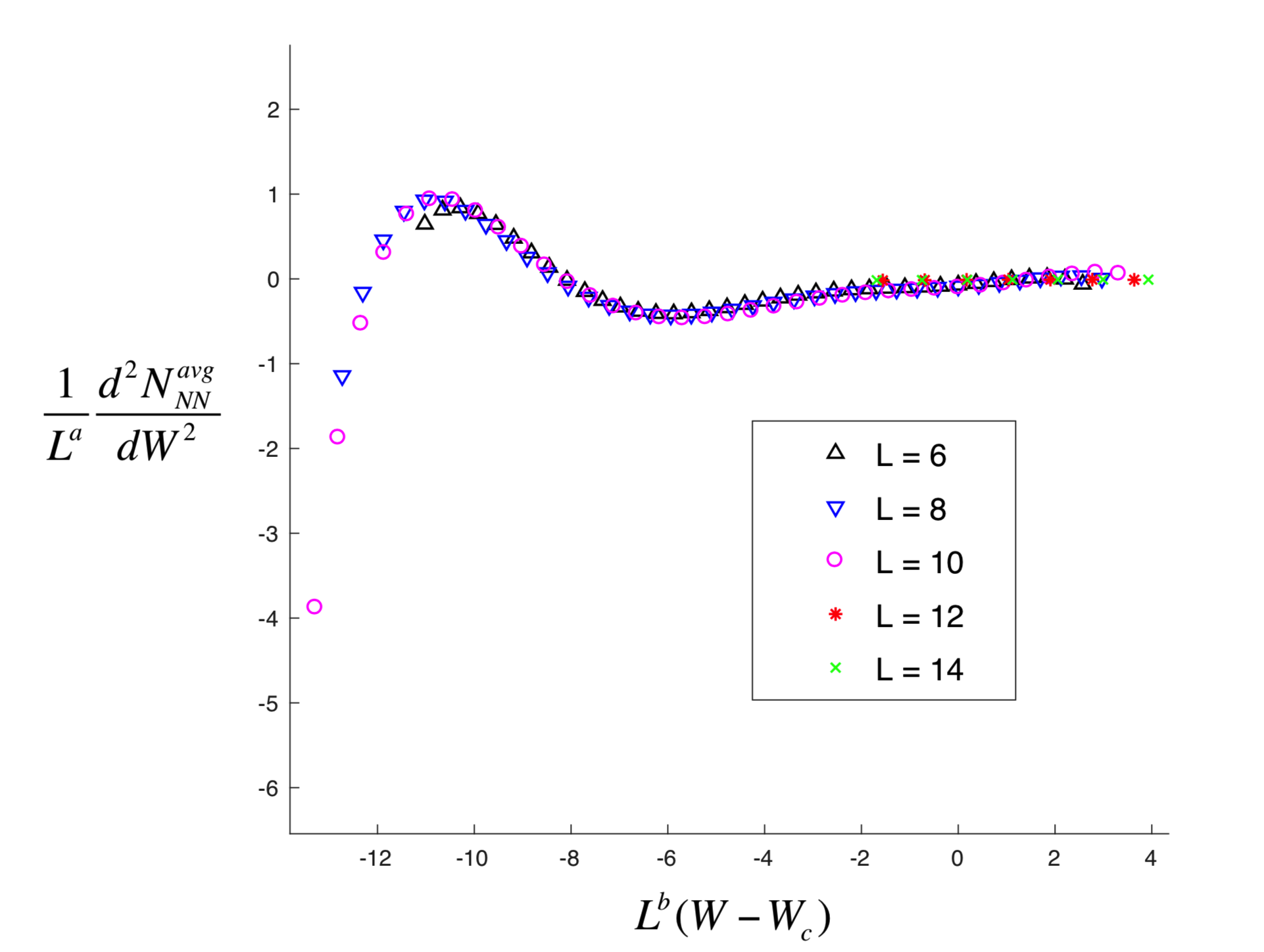}
\caption{\label{fig:Spin1NegCollapse} (Color online) Total nearest-neighbor negativity data in the spin-1 system is shown for various system sizes in relation to the scaling function in Eq.~\ref{eq:MBLCollapse}. The functional form is the same as for the spin-$1/2$ system considered above and in Ref.~\cite{Bera}; however, the fit parameters differ significantly. We find the best data collapse when $W_c = 4.7$, which is consistent with the estimates of $W_c$ that we obtain from the negativity itself, and from the decay in the negativity values as a function of the distance between sites.}
\end{figure}

 \begin{figure}
\includegraphics[width = 90mm]{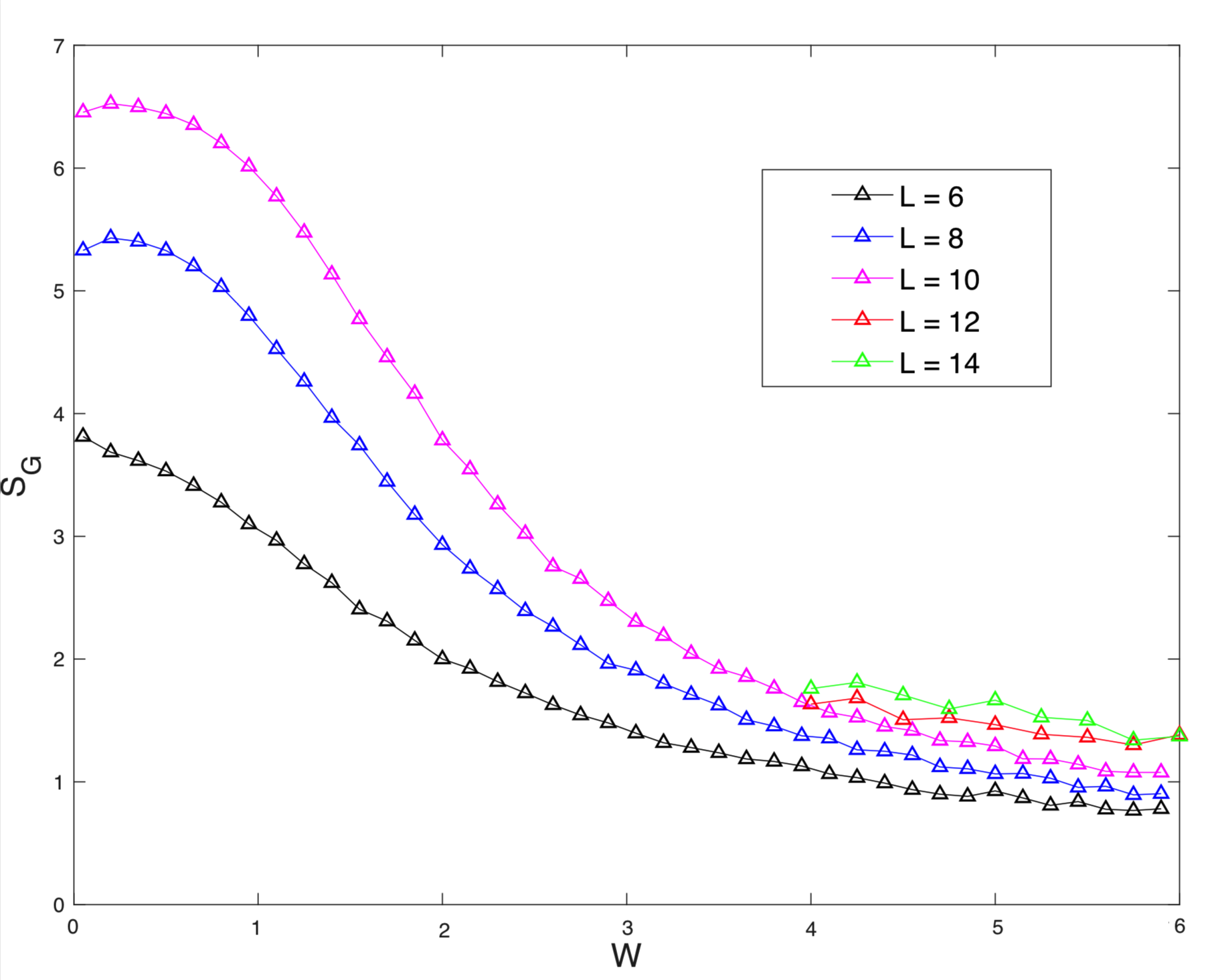}
\caption{\label{fig:Spin1GE} (Color online)  The geometric entanglement data for various system sizes in a disordered, spin-1 Heisenberg chain. This measure of global entanglement is seen to decrease as the disorder strength parameter $W$ increases, in the same manner seen for the spin-1/2 system considered above. The data are consistent with the onset of a localized phase for large $W$.}
\end{figure}

  \begin{figure}
\includegraphics[width = 90mm]{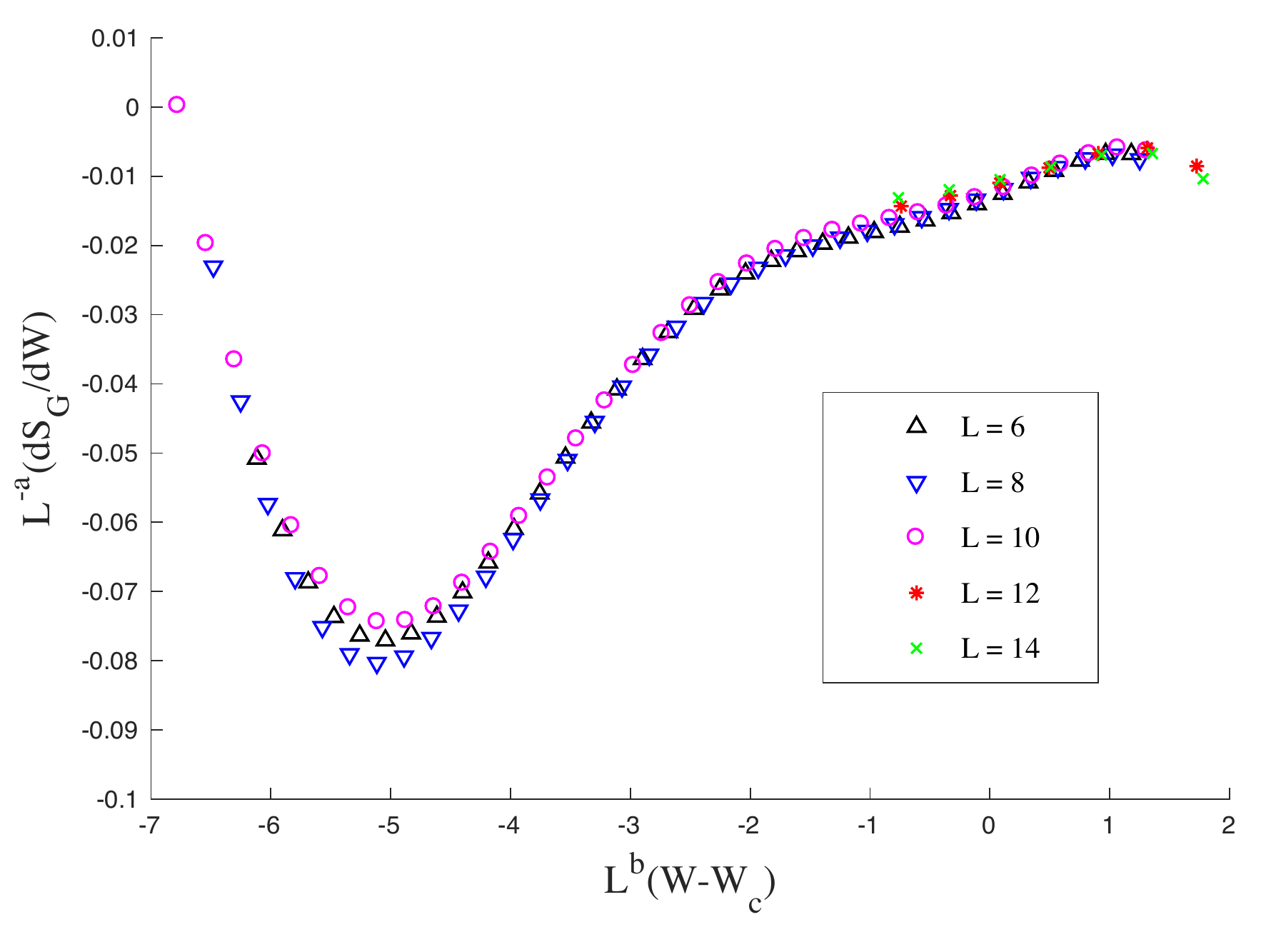}
\caption{\label{fig:Spin1GECollapse} (Color online) The geometric entanglement data from the random-field spin-1Heisenberg model, plotted for various system sizes in relation to the scaling function defined by Eq.~\ref{eq:GECollapse}, with $a = 1.2$, $b = 0.2$ and $W_c = 4.7$. Because of noisier data, weaker sensitivity to the choice of $W_c$ and greater sensitivity to the details of the fitting performed before computing the derivatives, this does not definitively identify $W_c = 4.7$ as the transition, as the collapse can still be seen for other choices of $W_c \in (4.5, 5)$. However, the presence of data collapse is consistent with our results from other methods, and supports the overall conclusion.}
\end{figure}

Of course, we can also turn to the geometric entanglement for further evidence. For the spin-1 system, we again see a similar pattern of global entanglement which decreases significantly as the disorder grows stronger (Fig,~\ref{fig:Spin1GE}). This pattern, too, is consistent with the appearance of a localized phase somewhere between $W_c = 4$ and $W_c = 5$. And while the greater statistical noise in our data for spin-1 geometric entanglement makes it challenging to use data collapse to definitively predict a transition point, we can at least show that it is not inconsistent with any of our other metrics. Fig~\ref{fig:Spin1GECollapse} shows the finite size scaling for the specific case of $W_c = 4.7$, for which we find relatively satisfying data collapse when $a = 1.2$ and $b = 0.2$, using the same functional form from Eq.~\ref{eq:GECollapse}. Though the overlap is not quite as complete as was the case for the spin-$1/2$ model, we note that the regions of greatest disagreement occur where the curvature is largest, precisely where our numerical fitting is the most sensitive and likely to deviate from the true underling entanglement curve.
It is important to note that for other values of $W_c$ in the range of 4.5 to 5 $W = 4$, it is also possible to achieve a comparable level of data collapse using slightly different parameters, and that because of the small number of points, the fit used in the $L= 12$ and $L= 14$ is somewhat sensitive to changes in $W_c$. Hence, this method alone is not \textit{predictive} of a transition at exactly $W_c = 4.7$, though it is certainly \textit{consistent} with this result as obtained from other methods. Because there is no \textit{ a priori} reason to expect data collapse of any kind, this is still good evidence that the transition exists and that it occurs in the neighborhood predicted by our other techniques.

\begin{figure}
\includegraphics[width = 90mm]{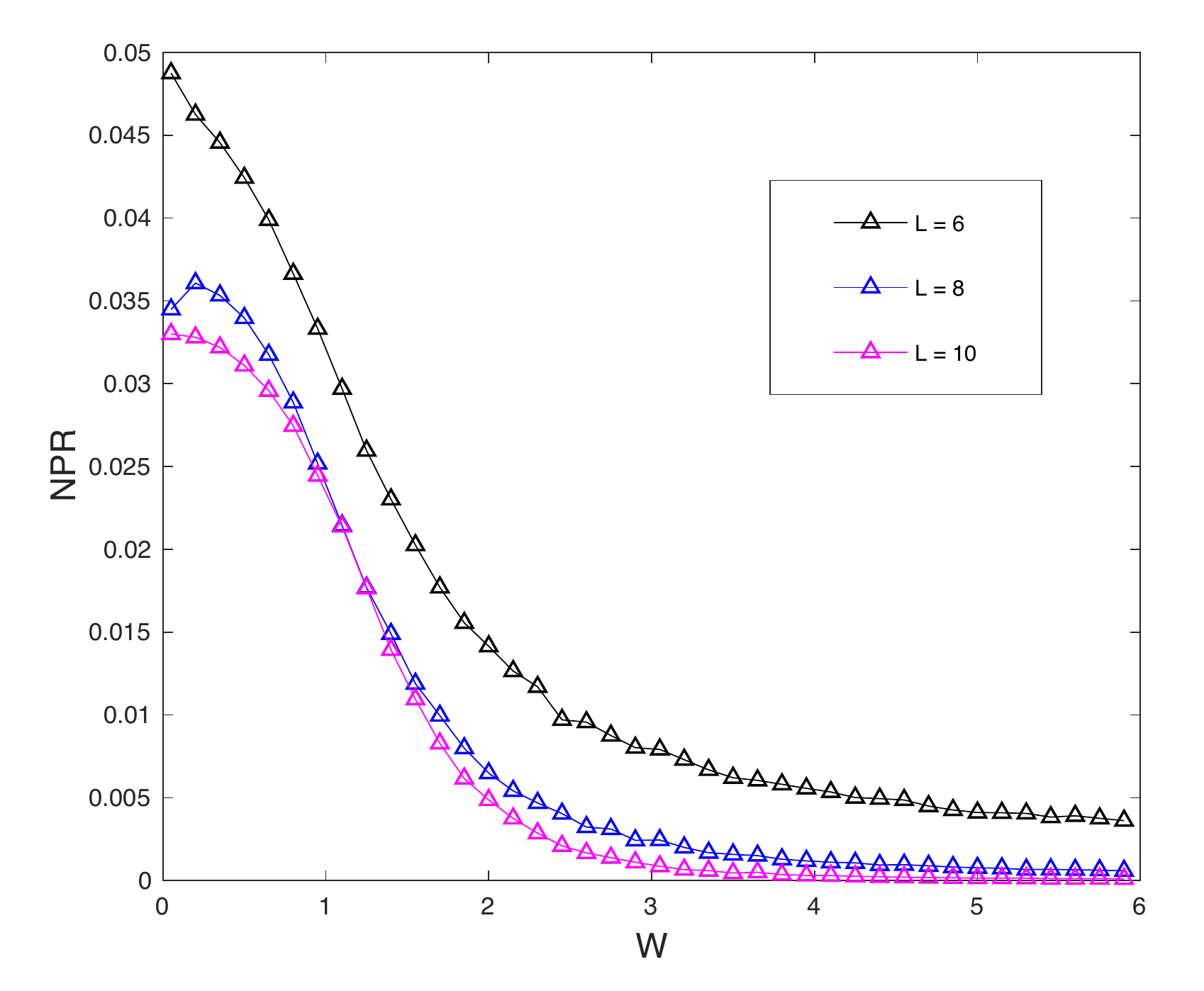}
\caption{\label{fig:Spin1NPR} The normalized participation ratios (NPR) for the disordered Heisenberg model, as a function of the strength of the disorder parameter. Though this quantity can be computed only for short, exactly-solvable systems, the general behavior, in which the NPR tends to vanish for larger systems and stronger disorder, is the same as that which was an indication of an MBL phase in the case of the spin-1/2 model considered above.}
\end{figure}

As a final method to demonstrate this localized-to-delocalized transition, we can consider also the behavior of the NPR, a localization indicator familiar from spin-1/2 systems. Though at present there is no efficient way to compute NPR from a large matrix product state, and we are therefore limited to the shorter system sizes, the behavior of this quantity even in this regime also supports the idea of an MBL phase for roughly $W \geq 4$. This is visible in Fig~\ref{fig:Spin1NPR}, since as the system length increases the NPR can be seen trending towards zero in this regime, in precisely the same manner as for the spin-1/2 system (c.f. Fig~\ref{fig:Neg}). 

Taken together, the relative agreement between these various entanglement methods is clear evidence that such a transition exists. Of course, as with any MBL study, it must be remembered that true localization is a phenomenon in the thermodynamic limit. Finding localization in a finite system is not by itself of particular interest; the key is to find evidence that this behavior might persist as the system size becomes infinite. The data above suggest that this is so for the spin-1 disordered Heisenberg model, as well as for the more widely-studied spin-1/2 case, but future studies reaching even longer system sizes will be necessary to give a more conclusive determination. 

\section{Summary}\label{Summary}
In this paper we have considered two measures of entanglement, the total nearest-neighbor concurrence and the geometric of entanglement, as indicators of MBL phenomena in spin chain systems. We build upon the work in Ref~\cite{Bera}, which considered the total nearest-neighbor concurrence, to show that both metrics can serve as indicators of localization. Furthermore, we demonstrate that by studying the finite-size scaling behavior for evidence of universal scaling, both metrics to estimate the critical disorder of an MBL system. To allow consideration of the longest possible spin chains, we study states produced by both exact diagonalization and by a numerical technique based on matrix product states~\cite{SIMPS}.

We first demonstrate the use of these metrics in the context of a disordered, spin-1/2 Heisenberg model, which is among the most widely-studied systems displaying evidence of MBL. Both measures of entanglement produce estimates of the critical disorder which agree with prior literature studying this model~\cite{Alet}. To further validate the behavior, we also compare to the normalized participation ratio, a long-established indicator of disordered behavior in its own right. 

The principle benefit of considering total nearest-neighbor negativity and geometric entanglement is that both measures can be computed for systems with a local dimension higher than two. To demonstrate this, we consider also the disordered spin-1 Heisenberg model, which to our knowledge has not previously been studied in detail as an example of a system displaying MBL. Our use of MPS algorithms here allows us to study system sizes up to $L = 14$. Applying the same techniques as we used in the spin-1/2 case, we demonstrate clear evidence for the existence of a localized regime, with a rough estimate of the critical disorder of $W_c = 4.7$. This result is consistent with both negativity and geometric entanglement, and also agrees with the normalized participation ratio for shorter chains. 
 
The authors are indebted to Soumya Bera for helpful conversations elaborating the techniques used in the Ref~\cite{Bera} study of entanglement in MBL systems, and to Sriram Ganesan and Josh Deutsch, for insightful feedback about the role of numerical simulations in studying quantum many-body systems. C.G.W and T.-C.W. acknowledge support from the National Science Foundation via Grants No. PHY 1333903 and No. PHY 1314748. T.-C.W. also acknowledges support from NSF grant No.PHY 1620252.
 
\bibliography{MBLPaper}{}
\bibliographystyle{unsrt}

\appendix
\section{SIMPS Convergence scheme}\label{AppendixSIMPSConvergence}
Recall that the SIMPS algorithm~\cite{SIMPS} consists of two nested loops, described in Sec.~\ref{Algorithms}. In our implementation, the numerical tolerance and the convergence in these two loops is controlled by five parameters, with two parameters for the ``inner" loop and three for the outer. 

In the inner loop (where we are sweeping across the state to find $\ket{\phi_{n+1}} = \tilde{H} \ket{\phi_n}$) we set our convergence based upon the quantity $\delta_1 \equiv |\bra{\phi_n}O\ket{\phi_{n+1}}|$, which clearly should converge to zero when $\ket{\phi_{n+1}}$ is in the desired state because $O = \tilde{H}^{-1}$. This is a natural choice as a convergence check because it can be computed for free~\cite{SIMPS}, since it is equivalent to the overlap $\innerproduct{A_{[j]}}{B_{eff}}$. Hence, we set a tolerance $\epsilon_1$ and consider the sweeping process to have converged when $1-\delta_1 < \epsilon_1$. 

In practice we have observed that in the early stages of the algorithm, this sweeping back and forth across the state often ``stalls" in the sense that $\delta_1$ may asymptotically approach some limit which is strictly less than 1, and hence that the convergence criterion with respect to $\delta_1$ may never be reached. This is likely due to finite bond-dimension effects: although we are searching for a final state $\ket{\phi_{n\to\infty}}$ which we believe can be efficiently represented by an MPS because of its localized properties, there is no clear reason why the intermediate states $\ket{\phi_{n}}$ must be similarly expressible. Hence the convergence may stall when the current MPS cannot represent the desired intermediate state $\tilde{H}\ket{\phi_{n}}$, but rather only the best approximation available at the given bond dimension. To handle this, we consider after each sweep the quantity $\delta_2 \equiv \delta^{(k)}_1 - \delta^{(k-1)}_1 $ which simply measures the change in our convergence quantity between successive sweeps $k-1$ and $k$. We set an accompanying tolerance $\epsilon_2$ and exit the loop, assuming the process to have stalled, if $\delta_2$ drops below this threshold. Typically, this kind of early exit is tolerable, and may actually improve the speed of the algorithm by avoiding fruitless sweeping steps. The resulting vector $\ket{\phi_{n+1}}$ will be imperfect, but will still have greater overlap with the target state than $\ket{\phi_{n}}$, and because of the nature of the power method being used, additional applications of $\tilde{H}$ in subsequent steps will continue to rotate us towards the desired state. In practice, we have seen that as long as the quantity $\delta_1$ is relatively close to 1 (e.g. perhaps $\delta_1 > 0.8$) the state will continue converging towards the desired target, at which point the MPS representation will become more accurate and the ``stalling" problem will vanish.

In the outer loop, we must decide how many times $\tilde{H}$ must be applied before the target state has been reached to a good approximation. We use two additional quantities to measure this. The first is simply $\delta_3 = |E_n - E_{n+1}|/E_n$, the relative change in the energy of the state (with respect to the \emph{original} Hamiltonian) between steps $n$ and $n+1$. We also examine the variance of the energy $\delta_4 = \langle \Delta H^2 \rangle$. This can be an important convergence check for any algorithm, but is particularly useful in this context, because it vanishes for an exact eigenstate. For our purposes, to understand the overall spectrum of the disordered Hamiltonians, we need to be sure we are studying proper localized eigenstates (rather than simply a superposition of such). Because of it's important, we set a relatively strict convergence threshold for this quantity ($1 \times 10^-7$), and reject any states whose variance is too large. 

Unfortunately, even with these checks in place, in practice we have occasionally seen the algorithm converge towards what appears to be a superposition of eigenstates (as judged by comparison to an exact diagonalization result). This is not entirely surprising, since in practice two closely neighboring states will both produce large eigenvalues in the shifted-and-inverted spectrum, and although theoretically in the large-$n$ limit only the dominant eigenvector will remain, it may take many steps before this is the case, and components from the neighboring state will persist longer than any other. In a superposition of such nearby states, there may be very little change in the energy values after successive steps of the algorithm ($\delta_3$), and even a relatively small energy variance ($\delta_4$). To avoid being fooled into accepting such a superposition, we use one additional convergence check: the overlap between states at subsequent steps of the algorithm, $\delta_5 = \innerproduct{\phi_n}{\phi_{n+1}}$. When two states are close to the target energy, this quantity will tend to decrease at first, but then may begin to increase temporarily until the algorithm fully settles in to one state. In practice this is a very strong convergence check.
%In practice this is likely to be the strongest convergence check; when it is close to 1, the properties of the state are almost certainly converged.

Naturally, one might wonder if the performance of the algorithm can be improved by simply squaring the MPO (i.e., taking $\tilde{H} \to \tilde{H}^2$) so that at each step of the outer loop we apply more than one inverse. Our tests with this method show that in practice, the required increase in the bond dimension of the MPO slows the construction of the structures $O_{eff}$ in a manner which overcomes the potential advantage. The sweeping procedure also experiences more instability and has greater difficulty converging to a result.

\end{document}